%% file: template_arxiv.tex
\documentclass{article}

\usepackage{arxiv}

\usepackage[utf8]{inputenc} % allow utf-8 input
\usepackage[T1]{fontenc}    % use 8-bit T1 fonts
\usepackage{hyperref}       % hyperlinks
\usepackage{url}            % simple URL typesetting
\usepackage{booktabs}       % professional-quality tables
\usepackage{amsfonts}       % blackboard math symbols
\usepackage{nicefrac}       % compact symbols for 1/2, etc.
\usepackage{microtype}      % microtypography
\usepackage{graphicx}
\usepackage{natbib}
\usepackage{doi}

\title{Review of real-time data processing for collider experiments}

\author{ 
V. V. Gligorov \\
LPNHE, Sorbonne Universit{\'e} \\
Paris Diderot Sorbonne Paris Cit{\'e} \\ 
CNRS/IN2P3, France \\
\href{mailto:vladimir.gligorov@cern.ch}{vladimir.gligorov@cern.ch} \\
\And 
V. Rekovi\'{c} \\
Vinca Institute of Nuclear Sciences \\
National Institute of the Republic of Serbia \\ 
University of Belgrade, M. Petrovica Alasa 12-14 \\ 
Belgrade, Serbia \\
\href{mailto:vladimir.rekovic@cern.ch}{vladimir.rekovic@cern.ch}
}

\renewcommand{\headeright}{}
\renewcommand{\undertitle}{}
\renewcommand{\shorttitle}{Review of real-time data processing for collider experiments}

\hypersetup{
pdftitle={A template for the arxiv style},
pdfsubject={q-bio.NC, q-bio.QM},
pdfauthor={David S.~Hippocampus, Elias D.~Striatum},
pdfkeywords={First keyword, Second keyword, More},
}

\begin{document}
\maketitle

\begin{abstract}
We review the status of, and prospects for, real-time data processing for collider experiments in experimental 
High Energy Physics. We discuss the historical evolution of data rates and volumes in the field and place them 
in the context of data in other scientific domains and commercial applications. We review the requirements for 
real-time processing of these data, and the constraints they impose on the computing architectures used for 
such processing. We describe the evolution of real-time processing over the past decades with a particular focus 
on the Large Hadron Collider experiments and their planned upgrades over the next decade. We then discuss how the 
scientific trends in the field and commercial trends in computing architectures may influence real-time processing 
over the coming decades.
\end{abstract}

\input{body}

\section*{Acknowledgements}
VVG acknowledges support by the European Research Council under Grant Agreement number 724777 ``RECEPT''. The authors 
would like to thank Alessandro Cerri (University of Sussex) for the kind permission to use Figure~\ref{fig:data_rates_cerri} in this article. 
We thank George Kour for preparing the arXiv template used for this preprint.

\bibliographystyle{unsrtnat}
\bibliography{my-bib-database}  %%% Uncomment this line and comment out the ``thebibliography'' section below to use the external .bib file (using bibtex) .

\end{document}

%% file: body.tex
\section{Introduction}
\label{intro}

Experimental high-energy physics (HEP) is characterised by tremendous quantities of data, whether considered in terms
of the instantaneous data rates produced by individual experiments or in terms of the overall data volumes analysed
when searching for particles or measuring their fundamental properties. Not only is it impractical to record
all data produced by HEP experiments, but it is even less practical to distribute these data to the hundreds or 
thousands of physicists who should then analyse it. This impracticality has made real-time data processing a central
element of nearly all HEP experiments. Perhaps surprisingly, considering that nearly all collider experiment publications 
use data that have undergone non-trivial real-time processing, there are few pedagogical reviews covering the general 
requirements and principles of HEP real-time processing systems. The closest we can find are 
Refs.~\cite{LINDENSTRUTH200448,6418180,Cittolin:2012zz}, none of which is available 
under open access or on arXiv. Of these the Lindenstruth and Kisel article is the most general but is also now almost 
twenty years old. This motivates our review, which focuses on collider experiments. We discuss the general constraints and 
objectives of HEP real-time processing systems, briefly recapitulate their history, and then take a more detailed look at 
the real-time processing systems of Large Hadron Collider experiments and their future upgrades. We conclude with some 
reflections on possible future trends in real-time processing in our domain. 

In this review we assume a basic undergraduate-level
familiarity with experimental HEP and the design of HEP detectors, but no specialist knowledge about computing. We
try to use words in their ``natural English meaning'' where possible, and define specialist terminology or jargon as 
it is introduced. Given how quickly computing architectures of all types are developing today, we deliberately stay away 
from statements and discussions which are specific to any given architecture unless necessary to illustrate a broader
conceptual point. If you are interested in learning about specific real-time processing architectures the best resource 
to consult is the relevant technical design report of the experiment in question -- indeed it is our hope that this 
review will help the reader to contextualise such specialist literature. Our choice of examples with which certain 
points are illustrated are of course biased by the experimental collaborations we have been members of, but this review 
does not pretend to act as a history of HEP real-time processing systems or a commentary on their relative importance 
to the field and our choice of examples should not be intepreted in such a light.

To underline from the beginning that real-time processing is not a novel problem to HEP, consider ALEPH, one of the four 
Large Electron Positron (LEP) collider experiments whose data taking began in the late 1980s. An unprocessed random 
ALEPH ``event''\footnote{For the purposes of this review an event designates one nominal unit of data used in physics 
analysis. In collider experiments it typically corresponds to the data produced during one crossing of the colliding
beams.} was around 30~kbytes in size, while an event containing a $Z^0$ boson was several hundreds of kbytes. 
As a crossing of the LEP particle beams occured every 22~$\mu$s, this meant a total unprocessed data rate of 
1.36~Gbytes per second. LEP ran for around $10^6$ seconds per year and consumer hard drives cost~\cite{prices} around 
$10^4$~dollars per GByte in 1989. Storing all data to either tape or disk was therefore out of the question even 
without taking into account the additional costs, sometimes up to an order of magnitude in overhead, of backing up 
the data and using hard disks which could operate continuously for many years without breaking. Transmitting this data 
would have been similarly impractical at a time when top of the line Ethernet connections had speeds in the hundreds 
of kBytes per second. Both the data rate and volume therefore had to be reduced by around five orders of magnitude to 
bring them into the realm of what could be stored and distributed.

This relationship between generated data rates, data volumes, and storage costs have remained rather similar between  
LEP and the Large Hadron Collider (LHC) collider built almost twenty years later. The LHC is a proton-proton collider 
in which beam crossings take place every 25~ns. Take as an example LHCb~\cite{Aaij:2014jba}, which has the smallest 
individual event sizes of the LHC's four ``main'' experiments. The LHCb detector produces around 30~kbytes of data 
for a randomly selected LHC beam crossing. This gives an overall data rate of around 1.2~Tbytes 
per second, around four orders of magnitude larger than the ALEPH zero-suppressed rate. The larger ATLAS and CMS 
detectors~\cite{Aad:1129811,Chatrchyan:1129810} have raw data rates which are around an order of magnitude larger still. 
On the other hand, consumer drives were roughly $0.1$~dollar per Gbyte when the LHC started up in 2009, 
around five orders of magnitude smaller than in 1989. Similarly typical Ethernet connections in 2009 had speeds of 
hundreds of Mbytes per second, around three orders of magnitude better than at the time of LEP. So the data still
has to be reduced by around four orders of magnitude through real-time processing.

\begin{figure}[t]
    \centering
    \includegraphics[width=0.7\textwidth]{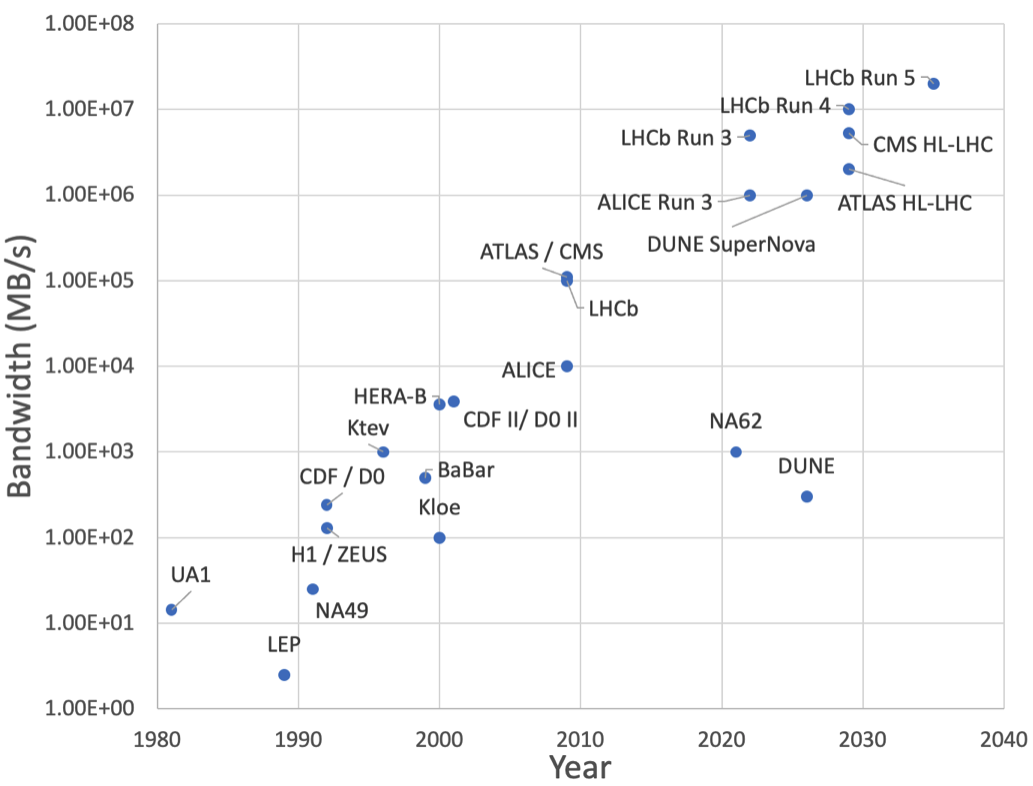}
    \caption{Instantaneous data rates (bandwidth) of HEP experiments over the past four decades. Data compiled by A. Cerri and used with permission.}
    \label{fig:data_rates_cerri}
\end{figure}

This near-logarithmic evolution of HEP experiment data rates is illustrated in Figure~\ref{fig:data_rates_cerri}.
It can also be put in the context of global internet traffic. ALEPH's 1.3~Gbyte of data per second 
dwarfed the 1~Tbyte of data per month being transmitted in 1990 on networks which preceeded the internet. By the time 
of the LHC in 2009 worldwide internet data rates had grown to around 15~Ebytes per month, which at around 5.8~Tbytes 
per second was still smaller than the instantaneous data rates of the ATLAS and CMS detectors, though a few factors larger 
than that of LHCb. It was not until around 2014-2015 that global internet dataflow finally outgrew the data rates 
produced by the LHC experiments. 

A non-trivial real-time processing of data is therefore necessary for almost all HEP experiments. This processing is
typically referred to as ``triggering'' in the jargon of the field, a historical term which dates back to the period 
when all such processing was implemented in custom electronics and issued a ``trigger'' to read out the full detector 
when a pattern of interest was found. Any non-trivial real-time processing will necessarily and irreversibly distort 
or bias the data by preferring certain types of events (and by discarding others). It is crucial that such distortions
are both as small as possible and as comprehensible as possible. Specifically the data must not be biased in ways which
make later physics analysis impossible, for example by inducing false structures mimicking the presence of new particles. 
They must also not be biased in ways which induce limiting systematic uncertainties on the precision with which physical 
quantities of interest can be inferred from the data. We begin our review of how well-designed real-time processing 
architectures achieve their goals with a closer look at the boundary conditions within which they operate.

\section{Constraints and requirements on real-time data processing}
\label{sec:rtaboundaries}

\begin{figure}[t]
    \centering
    \includegraphics[width=0.48\textwidth]{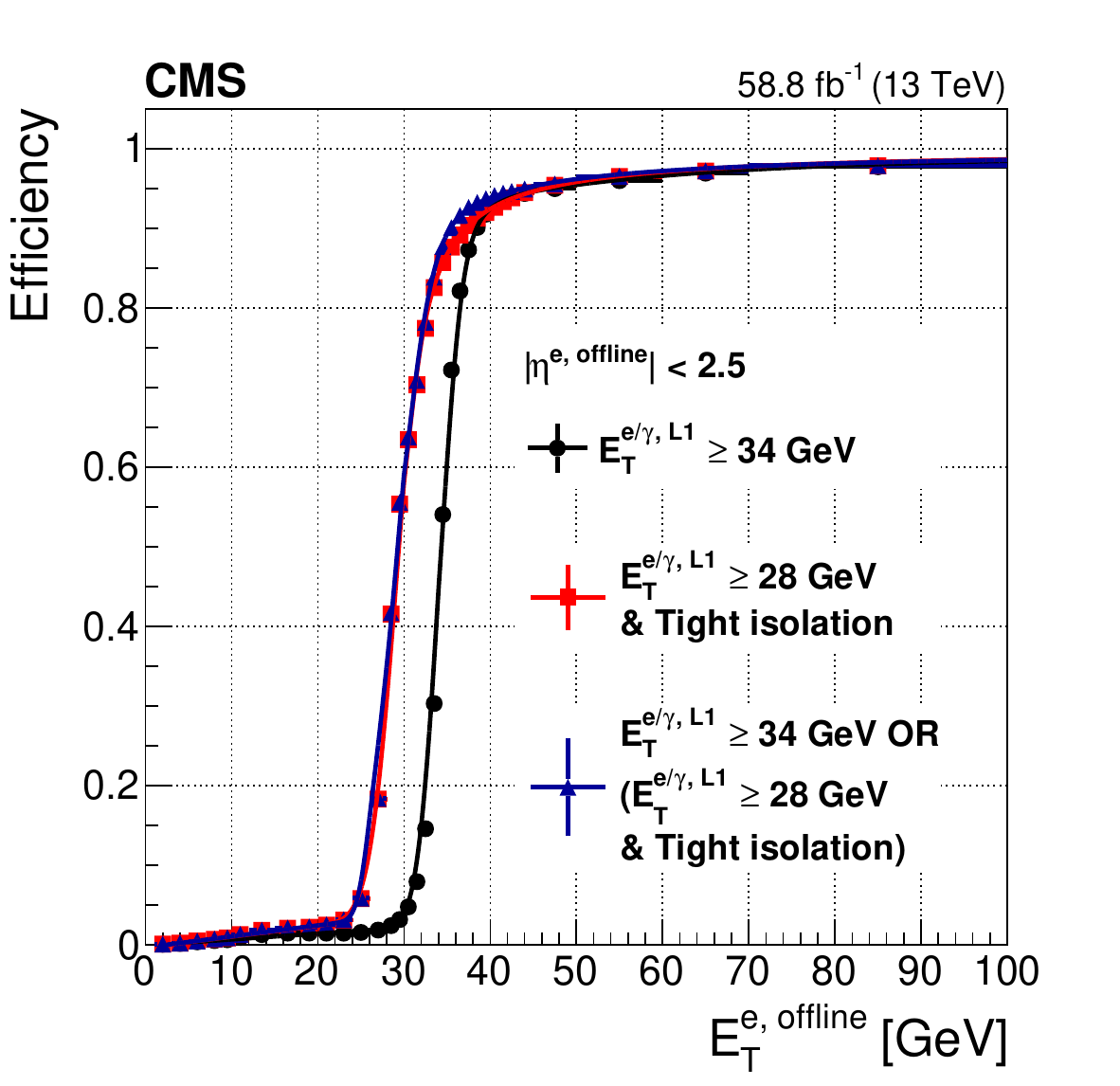}
    \includegraphics[width=0.48\textwidth]{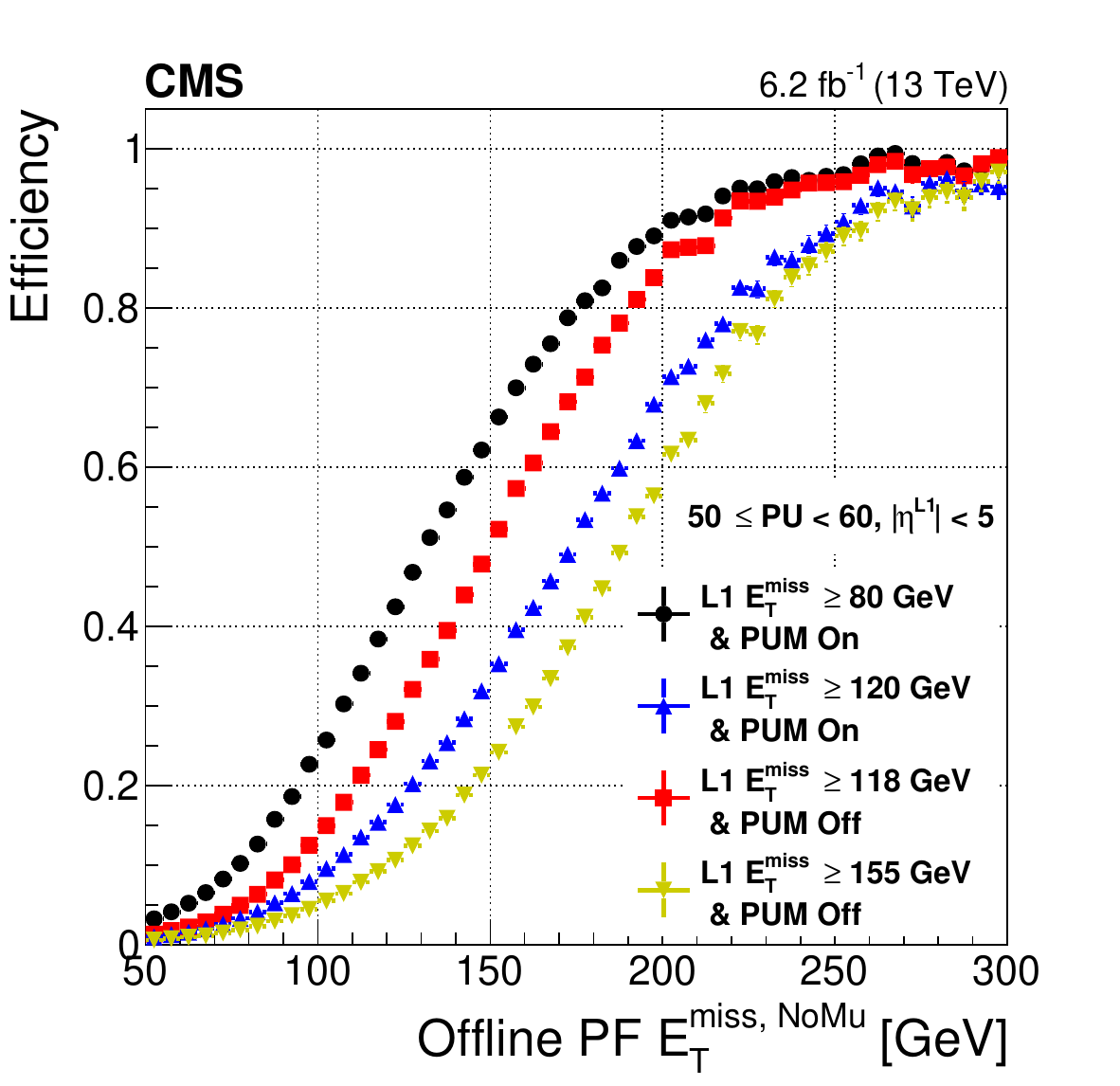}
    \caption{A comparison of the efficiency of the CMS first-level (left) electromagnetic calorimeter and 
    (right) missing transverse energy triggers as a function of the transverse momentum of the offline 
    physics analysis objects associated with the trigger decision. In the left case the ``turn-on'' curve is sharp
    and essentially reaches the trigger's efficiency plateau before the requirement which will be applied in offline 
    analysis, so that the trigger efficiency can be modelled as a single number. In the right case the shape of
    the turn-on curve must be modelled as well, with candidates which will fail the offline requirement passing the 
    trigger criteria and candidates which would pass offline requirements failing it. It is clearly desirable to design 
    a trigger as in the left case when possible. Reproduced from~\cite{CMS:2020cmk}. }
    \label{fig:turn-on-curves}
\end{figure}

In high-energy physics real-time, or near-real-time, processing of data which happens before the data is recorded to 
permanent storage is typically called ``online'', while any processing of data which has already been recorded 
to permanent storage is called ``offline''. The principal requirement which drives the design of any online HEP 
processing is the amount by which it must reduce its incoming data rate before the data are sent offline. As mentioned 
in the introduction this requirement is driven not only by the cost of storing the data, but equally by the cost of 
making the data available for offline physics analysis. A set of secondary requirements ensure that the selected data 
is usable for physics analysis. The online processing must be reproducible in simulation, and it has historically 
also been considered desirable that it is deterministic and reproducible at the bit level if the same data 
are processed multiple times. While statistical reproducibility\footnote{In practice this means that any deviations from 
perfect reproducibility should be small compared to other data-simulation differences and systematic uncertainties associated 
with the relevant physics analyses.} in simulation is a strong 
requirement, bit-level reproducibility in data is generally impractical with highly parallel computing architectures 
and is rarely of critical importance for physics analysis. The online processing must annotate the raw detector data 
with suitable metadata and provenance information so that a physics analyst can deduce how any given data file was 
processed. In particular information about derived quantities and objects that were created during the online processing
and led to the decision to record a given chunk of data to permanent storage must be recorded with the data itself.
Finally the online processing must be robust against reasonable variations in the underlying detector performance,
and in particular against few-percent-level degradations in detector efficiency or the availability of electronic 
channels which typically occur in any detector as it undergoes radiation damage over a number of years.
It is also highly desirable that the quantities used to make decisions online correspond as closely as 
possible to the quantities used in offline physics analysis, as illustrated in Figure~\ref{fig:turn-on-curves}. This
is not however a strict requirement since in many cases technical constraints make it impossible to achieve a close
correspondence between these quantities.

While the requirements on real-time processing are mostly universal across experiments, the constraints which
drive the design of specific real-time architectures are more closely linked to specific experiments and facilities.
First of all, the incoming data rate may be delivered as a continuous stream or in discrete packets arriving at uniform
time intervals. The latter is common for collider experiments such as those at the LHC, while the former can be found
in fixed-target or neutrino experiments such as NA62~\cite{NA62:2017rwk} and DUNE~\cite{DUNE:2020lwj}. If the data 
arrives in discrete packets, typically called ``events'' in the jargon of the field, the constraints differ depending 
on the size of an individual event and the rate in Hertz at which the events arrive for processing. Among the LHC 
experiments this is illustrated by the LHCb and ALICE trigger systems which sit at opposite corners of this two-dimensional 
space: LHCb processing 100~kByte events at 30~MHz~\cite{LHCb:2023hlw} and ALICE processing 22~MByte events at 
50~kHz~\cite{ALICEO2}. While the overall data volume treated by the two experiments is fairly similar, the implications 
for how events are processed are very different, particularly when deploying computing architectures where the main 
limitation comes from memory rather than compute power. 

A second set of constraints concerns the physical design of the detectors and the available bandwidth for
transferring the data. The innermost layers of hermetic detectors, the most common design for collider environments, are 
typically used for reconstructing, or ``tracking'' charged particle trajectories. Reading out these layers requires 
placing material inside the detector acceptance, which induces additional multiple scattering and energy loss and 
therefore impacts on the detector resolution. The greater the rate at which the detector is to be read out, the greater 
the number of cables and hence material introduced into the acceptance. If this constraint means that the innermost 
layers cannot be read out at the full event rate without fundamentally limiting the physics performance, the 
experiment must operate a first-level trigger based on information from its outer layers. And since this first-level 
trigger must communicate with the readout electronics of the innermost layers to tell them when an event needs 
to be read out, it must operate at a fixed latency. This is why most such experiments, of which ATLAS and CMS are 
prominent examples, must operate fixed-latency first-level\footnote{As we discuss in 
Sections~\ref{sec:rtaprelhc}~and~\ref{sec:rtalhcrun1}, several notable past experiments used multiple levels of 
fixed-latency triggers, but modern experiments which use fixed-latency triggers almost universally use only one level.} 
triggers, which are typically implemented using custom ASICs or, increasingly, off-the-shelf FPGA processors embedded 
in custom electronics boards which handle the I/O. This is not the only reason why experiments use fixed-latency triggers, 
nor the only example of how a detector's geometry constrains its data processing, but it's the most generally illustrative one.

One concrete example of how physical detector constraints influence the design of real-time processing systems are the 
tracking triggers~\cite{CERN-LHCC-2017-020,CERN-LHCC-2020-004} of the upgraded ATLAS and CMS detectors, due to go online 
during the High-Luminosity phase of 
the LHC (HL-LHC) after 2029. As discussed in Section~\ref{sec:rtalhcrun1}, the original ATLAS and CMS detectors were 
not designed to use their tracking detectors at the first trigger level. This was predominantly due to the fact that the 
outer parts of detector (Calorimetry and Muon detectors) were able to provide sufficient discriminating 
power against background for the majority of the core physics program. On top of that the computing power which could be 
deployed in the electronics boards available at the time the detectors were being designed was limited, so that a prohibitive 
number of such boards would have been needed to cope with the complexity of the events which had to be processed. 
For these reasons the use of tracker information in the trigger was postponed to later trigger stages.

\begin{figure}[t]
    \centering
    \includegraphics[width=0.9\textwidth]{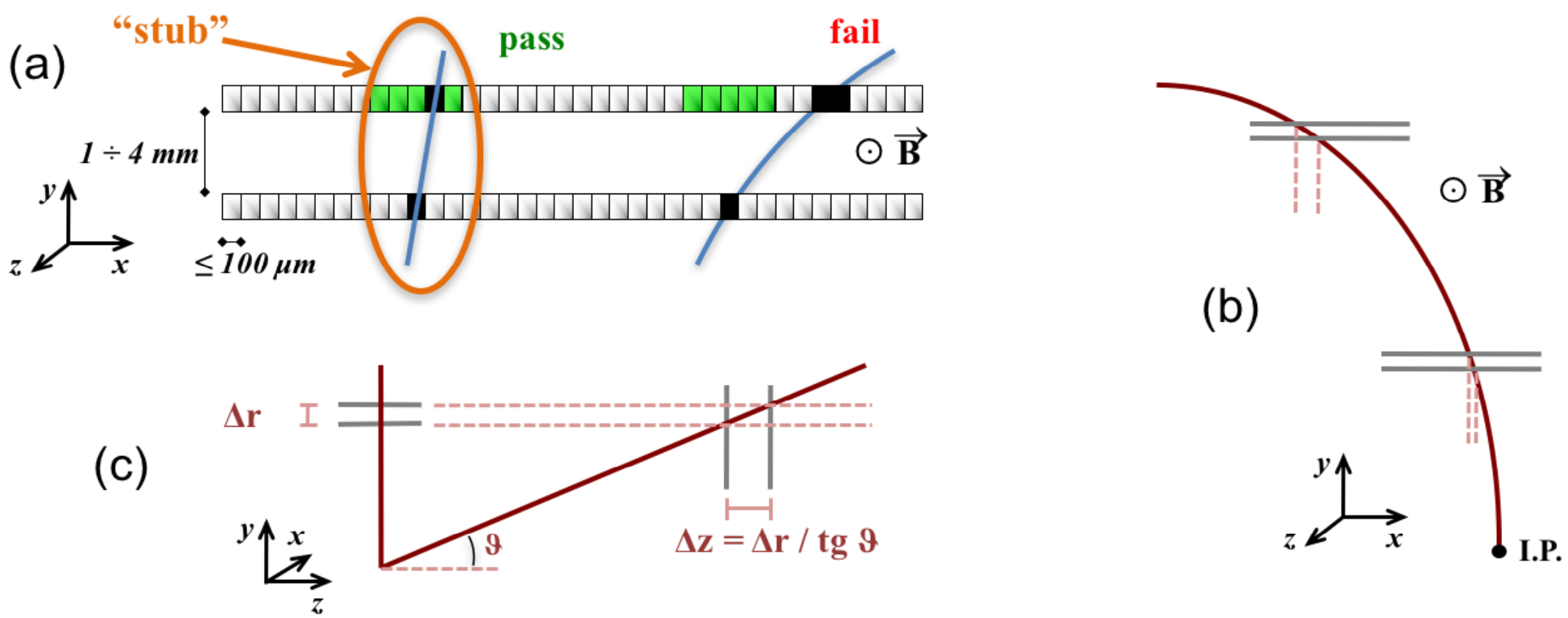}
    \caption{Illustration of the module concept which enables the CMS first-level tracking trigger. (a) Correlation of signals in closely-spaced sensors enables rejection of low transverse momentum particles; the channels shown in green represent the selection window to define an accepted stub. (b) The same transverse momentum corresponds to a larger distance between the two signals at large radii for a given sensor spacing. (c) For the end- cap discs, a larger spacing between the sensors is needed to achieve the same discriminating power as in the barrel at the same radius. Caption and Figure reproduced from~\cite{CERN-LHCC-2020-004}.}
    \label{fig:cmsl1tracktrigger}
\end{figure}

For the upgraded LHC collider, the HL-LHC, the instantaneous luminosity will be increased such that individual bunch crossings 
will contain up to 200 proton-proton collisions. This in turn means much larger QCD backgrounds which are difficult to distinguish 
with only calorimeter and muon system information. Therefore both ATLAS and CMS have rethought their trigger designs in order to 
introduce tracking information at earlier stages. In both cases it remains impossible to read out the tracking detectors closest 
to the beampipe at the full crossing rate, as described earlier. However CMS has placed the layers of its outer, silicon-sensor, 
tracking system in pairs which can then be read out into a set of custom electronics boards at the full crossing rate. This layout is 
illustrated in Figure~\ref{fig:cmsl1tracktrigger}. Under the 
assumption that particles come from the known proton-proton interaction region, the location of their hits in each pair of layers 
allows a simplified determination of their trajectory and therefore transverse momentum. This in turn allows the data rate to 
be greatly reduced before these proto-track candidates are extrapolated to other layers and used in the trigger decision. 
This interplay of detector geometry and real-time processing will give the upgraded CMS detector on-layer local stub reconstruction 
capabilities in the region $|\eta| < 2.4$ for particles with transverse momentum above 2~GeV, significantly improving the ability 
of the trigger system to identify charged particles and improving its kinematic resolution to nearly the levels previously 
achieved in only later, so called ``high-level trigger'' (HLT) stages. 

%This feature designed to discard particles with very low transverse momentum (below 2 GeV), together with  maximal allowed 
%latency for triggering increased from 3.2 to 12.5 $\mu$s will allow Phase-2 upgraded CMS Level-1 Trigger to reconstruct 
%Tracker tracks in hardware and use this important information for significantly improved particle identification and 
%reconstruction of kinematics, at the level of resolution previously possible only in the HLT.
%  $|\eta| < 4.0$, This new detector located closest to the collision point and composed of layers of high 
% granularity silicon sensor with coverage of
%For the first time in history of particle detectors on hadron colliders the first-level triggering will include tracks 
%reconstructed using hits recorded by the Phase-2 upgraded CMS Tracker.
%Current and past particle detectors at hadron colliders which are notorious for high particle content in collision events 
%have  The other reason was the insufficient computing power of 
%available processors in the custom-based electronic boards to be able to cope in sufficiently short time with the large 
%volumes of data from the most inner detectors, like Tracker, where the detector resolution and occupancy are highest.
%For that reason the use of Tracker information in the triggering was postponed to the HLT where the typical allowed latency 
%is on the order of tens or hundreds of miliseconds. 

HEP detectors typically transfer data which passes the first-level trigger selection over optical links into a custom data 
centre where it is further processed in real or near-real time. These optical links must be radiation-hard on the detector 
side, which places constraints on their design and, depending on the protocol used, can place constraints on how far from 
the detector the data can be reliably transferred. Since this review focuses on triggering rather than electronics and 
data acquisition, we do not cover this topic further. % but interested readers can find a comprehensive review in Ref.~\cite{}. 
The design of this data centre is itself subject to constraints based on the available physical space, power, and cooling 
capacity, which can significantly influence the types of computing architectures used. Typically, however, there are no 
longer any strict latency requirements for HLT systems. 

In recent decades the most common architecture 
used for high-level triggers have been CPUs, in particular x86 CPUs. Prior to roughly the mid-2010s these were almost 
universally used as serial processors, working on one event or one chunk of data at a time. Therefore the available 
computing power could be roughly translated into a constraint on the number of seconds which a single CPU (or single CPU 
core) could take to process one event or one chunk of data. While this did not perfectly translate to performance in the
data centre it was good enough that physicists developing trigger software generally did not have to worry about 
replicating the exact data centre environment when benchmarking their code. Modern CPUs however contain so many cores 
which share memory that their computational performance is often memory rather than compute-bound, and the real-world 
performance is consequently much more closely linked to the precise data centre environment. The constraint on available
computation performance is therefore more usefully expressed in terms of the number of events which a single server
must process each second when under full load. In addition today's data rates are sufficiently high that a further 
constraint arises from the frequency at which data can be copied in and out of the main processor memory without causing 
a bottleneck in the performance. All these considerations are even more true of GPU and other parallel architectures
than they are of CPUs. For this reason modern parallel computing architectures require dedicated testbench setups to 
measure and monitoring their performance, with all the associated maintenance overhead which this entails.

A final set of constraints comes from the physics program of the experiment and concerns the number of distinct ways
in which the trigger may make the decision that an event or chunk of data should be recorded to permanent storage. These
are often called ``trigger paths'', ``trigger menus'', or ``selections'' depending on the jargon of the experiment in
question. While all non-trivial trigger selections make their decision based on the properties of physics objects
reconstructed from the underlying detector data, developing, testing, deploying, and managing the trigger code is very
different if dealing with tens or thousands of selections. In addition to physics selections triggers must implement
a variety of technical selections which record data for calibration purposes, whether that is to calibrate the detector
performance, the luminosity recorded by the detector, or simply to randomly sample events in order to perform studies
of biases induced by the trigger itself. Among these, high-precision measurements of the recorded luminosity can lead 
to particularly stringent constraints on the stability of the data acquisition and real-time processing, as discussed
in Refs.~\cite{Dam:2021sdj}.

\section{Real-time data processing prior to the Large Hadron Collider}
\label{sec:rtaprelhc}

While CERN's intersecting storage rings established the principles behind hadron colliders instrumented by hermetic 
detectors with near~$4\pi$ solid angle coverage, it was the UA1 experiment at CERN's SPS collider that defined the 
basic architecture which most hadron collider triggers have followed over the past four decades. The trigger system 
of the UA1 and UA2 experiments is described in Ref.~\cite{Dorenbosch:1985cx}. While the UA2 trigger system was mainly 
based around a reconstruction of the detector's calorimeter in custom electronics, the UA1 trigger was more complex. 

The first and second level UA1 triggers were based on muon and calorimeter information, as the central drift chamber 
could not be read out at the full collision rate. Their task was to reduce the input rate of 250~kHz to around 20~Hz, 
with a maximum allowed output rate of 40~Hz determined by the roughly 25~ms time it took to read out UA1's central drift 
chamber and compress its data stream. Once an event was accepted by the second level trigger the entire detector would be 
read out and a third level trigger performed a full event reconstruction and the final rate reduction. The first 
level was implemented in custom electronics, and its output was buffered for 200~$\mu$s while waiting for the second 
level trigger decision. This second level trigger was implemented using 68020~CPUs which communicated via a custom 
readout bus with the first-level trigger, while a 68010~CPU handled transmission of accepted events to the third level. 
The third-level trigger was implemented using six 3081 emulators and reduced the event rate to the final target of 
$\sim\! 5$~Hz. The primary physics objective of the trigger system was to select events containing large transverse 
energy signatures from the decay of heavy particles, most notably $W$ and $Z$ bosons, and reject the majority of 
events containing only light QCD processes. The first and second level triggers achieved this objective by using 
muon and calorimeter system information, while the third-level trigger refined the first and second level decisions 
using tracking information from UA1's central drift chamber.

Multiple concepts used in the UA1 trigger have formed the foundation of later hadron collider trigger systems. The use 
of lookup tables to reconstruct hit patterns in the first-level muon trigger and the summing of neighbouring calorimeter 
cells to form first-level trigger clusters are both in use to this day. The sorting of muon hits by wire number while 
transmitting data to the second-level trigger is an early example of a very common technique to both reduce the 
required number of memory accesses and speed up computation in  pattern recognition algorithms. The use of offline-quality 
calibration constants in the second-level calorimeter trigger helps to align trigger decisions to offline criteria and 
is also commonplace today. The use of 
drift chamber information at the third trigger level to ``confirm'' the muon tracks and calorimeter objects found in 
the first two trigger levels is a specific example of a general technique which became commonplace in the field. 
The benefit of one trigger stage confirming the decision of the previous stage is that it simplifies the calculation 
of trigger efficiencies. In the ideal case there is a single underlying physics object of interest which is responsible 
for the trigger decision, and each trigger stage simply reconstructs this object with increasing precision and 
sophistication. Finally the use of the 3081 emulators for offline data processing outside beam time foreshadowed the 
way that today's collider experiments use their trigger systems as general-purpose data centres. 

In contrast to the SPS detectors, whose triggers sought a small fraction of heavy particle decays in a sea of 
light QCD, the LEP $e^+e^-$ collider detector triggers needed to collect all genuine $e^+e^-$ interactions while 
rejecting bunch crossings containing only beam backgrounds. While the required reduction in event rate was around 
five orders of magnitude in both cases, the physics content of the LEP detector triggers was consequently quite 
different than that of the SPS triggers. The ALEPH and DELPHI triggers are described in 
Refs.~\cite{decamp:in2p3-00005366,Bocci:274052} and follow the same basic architecture as already seen at UA1, with 
a three or four stage trigger in which the first two stages are implemented in custom electronics and decide whether 
the full detector should be read out, while the third (and fourth) stage is implemented in a batch of commodity 
processors and decides whether the event should be recorded to permanent storage. Characteristically for $e^+e^-$ 
collisions the trigger systems were essentially 100\% efficient for all physics signatures of interest, as well as for 
Bhabba scattering processes essential for properly calibrating the luminosity, which largely 
eliminated the need for modelling the trigger efficiency in physics analyses. Trigger designs at other $e^+e^-$ colliders, 
notably the $b$-factories BaBar and Belle (as well as Belle~II), have the same basic objectives.  

At hadron and lepton-hadron colliders, a trigger system's main job is to distinguish signatures of interest from 
QCD backgrounds. Since QCD backgrounds fall off exponentially with the transverse momentum or energy of the reconstructed 
signature, these quantities efficiently separate heavy signatures from QCD. Experiments may however also be interested 
in lighter signatures, for example from the decays of beauty or charm hadrons, tau leptons, or other particles whose 
mass is close to the general scale of QCD processes produced at the given collision energy. In this case the trigger 
must use the topology of the signature of interest to distinguish it from backgrounds. While muon stations and calorimeters 
can provide some information about event topologies, tracking information, in particular close to the collision point, 
is vital for efficient topological discrimination. The H1 and CDF trigger systems, described in 
Refs.~\cite{H1:1996prr,Baird:2001xc,CDF:1987dye,CDF:2003mka}, made vital steps towards introducing tracking information 
at earlier trigger stages. The H1 fast track trigger was implemented in a system of FPGA boards and used lookup tables 
to compare hit patterns in data with valid hit-pattern masks from simulated events. The CDF tracking trigger consisted 
of two components: the XFT which used outer layers of the tracking system to measure track transverse momenta, and the 
SVT which used inner tracking layers to measure the displacement of these tracks from the primary $p-\bar{p}$ interaction. 
The SVT in particular was crucial in order to distinguish beauty and charm hadron decays, with their characteristic 
displacement from the primary interaction, from light QCD backgrounds. The CDF trigger would combine a pair of 
displaced tracks in the SVT into a vertex and require that these tracks and the vertex itself satisfied minimum transverse 
momentum requirements. This basic signature continues to be the foundation of heavy flavour triggers at hadron colliders.

\section{Real-time processing architectures of the original LHC experiments}
\label{sec:rtalhcrun1}

ATLAS and CMS were designed as general purpose detectors which would be able to address same set of physics questions at the LHC. 
This defined a very broad physics program for these two experiments, including precision measurements of the Electroweak sector, 
the search for a Standard Model-like Higgs boson in the mass range between 115 and 200~GeV, as well as the broadest possible 
capabilities in terms of searching for beyond Standard Model particles and processes. In practical terms, the two detectors 
needed to process events from LHC collisions occurring at a rate of up to 40~MHz and retain on the order of 1~kHz of events 
containing the most interesting physics signatures such as high transverse momentum leptons, jets, or large quantities of missing 
transverse energy. It was also however crucially important to make the design of this real-time processing as flexible as possible, 
in order to allow the experiments to adapt their physics programs to unexpected discoveries and evolve them with time more generally.

\begin{figure}[t]
    \centering
    \includegraphics[width=0.9\textwidth]{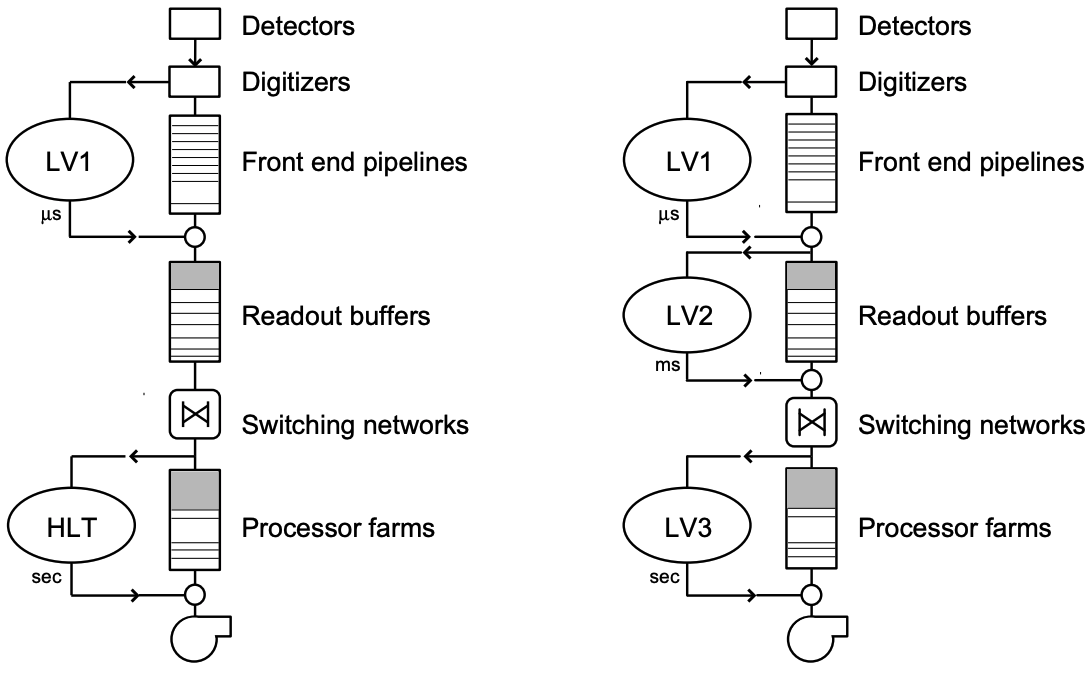}
    \caption{Data flow in the Trigger/DAQ system. Left: the CMS choice, with a single entity, the High-Level Trigger (HLT) providing all filtering after the Level-1 decision. Right: the HLT function is split into two stages (LV2 and LV3). Caption and Figure reproduced from~\cite{Cittolin:578006}. Our note: the rightmost diagram is largely representative of the ATLAS choice.}
    \label{fig:cmsandatlastriggers}
\end{figure}

The original ATLAS and CMS real-time processing systems are described in detail in 
Refs.~\cite{CERN-LHCC-98-014,Jenni:616089,Bayatyan:706847,Cittolin:578006}. As in many other areas the two detectors made different 
technological choices, which also improved the global robustness of the LHC physics program in case some of these choices turned out 
to be less than optimal. ATLAS opted for a then more traditional three-level system which closely followed the UA1 architecture. 
The first level was a fixed-latency muon and calorimeter trigger which reduced the event rate to around 70~kHz, while the second and 
third levels were implemented in a farm of commodity processors. The second level performed a full-granularity reconstruction in 
regions of interest defined by objects reconstructed at the first level, while the third level built the full granularity event 
information from the whole detector and performed an offline-like reconstruction and event selection. By contrast, CMS opted for a 
two-level approach. The first level was, like in ATLAS, based on muon and calorimeter information and was designed to select around 
100~kHz of events. After that, however, the full granularity events would be built using information from the whole detector and 
passed to the second-level trigger implemented in a data centre based on commodity commercial processors. The difference between these 
systems is summarized in Figure~\ref{fig:cmsandatlastriggers}. 

The CMS system was designed to be conceptually simpler, chiefly at the cost of the assumption that networking technology would evolve at sufficient 
speed to make the full detector readout possible at 100~kHz by the time datataking started. This, as it turns out, is a bet that 
worked out very well for CMS, although it should be emphasised that the ATLAS trigger design did not impose any fundamental limitations 
on the collaboration's physics program either. Both experiments were able to adapt their triggers to novel beyond SM signatures and 
retained sufficient flexibility to evolve their trigger systems towards performing analysis-like tasks in real time, as discussed 
in the next section. Nevertheless it is undeniably true that network bandwidth evolved, and continues to 
evolve, more quickly than computing power or affordable long-term data storage. Both the CMS and ATLAS HL-LHC trigger systems will 
consequently follow a two-level design, although the specific implementations continue to have differences. 
More generally speaking, as we discuss later in the context of Figure~\ref{fig:generaltriggerarch}, the choice for collider trigger 
systems today is essentially between a full detector readout without any fixed-latency trigger system, or for a single fixed latency 
level followed by a full detector readout. 

\begin{figure}[t]
    \centering
    \includegraphics[width=0.7\textwidth]{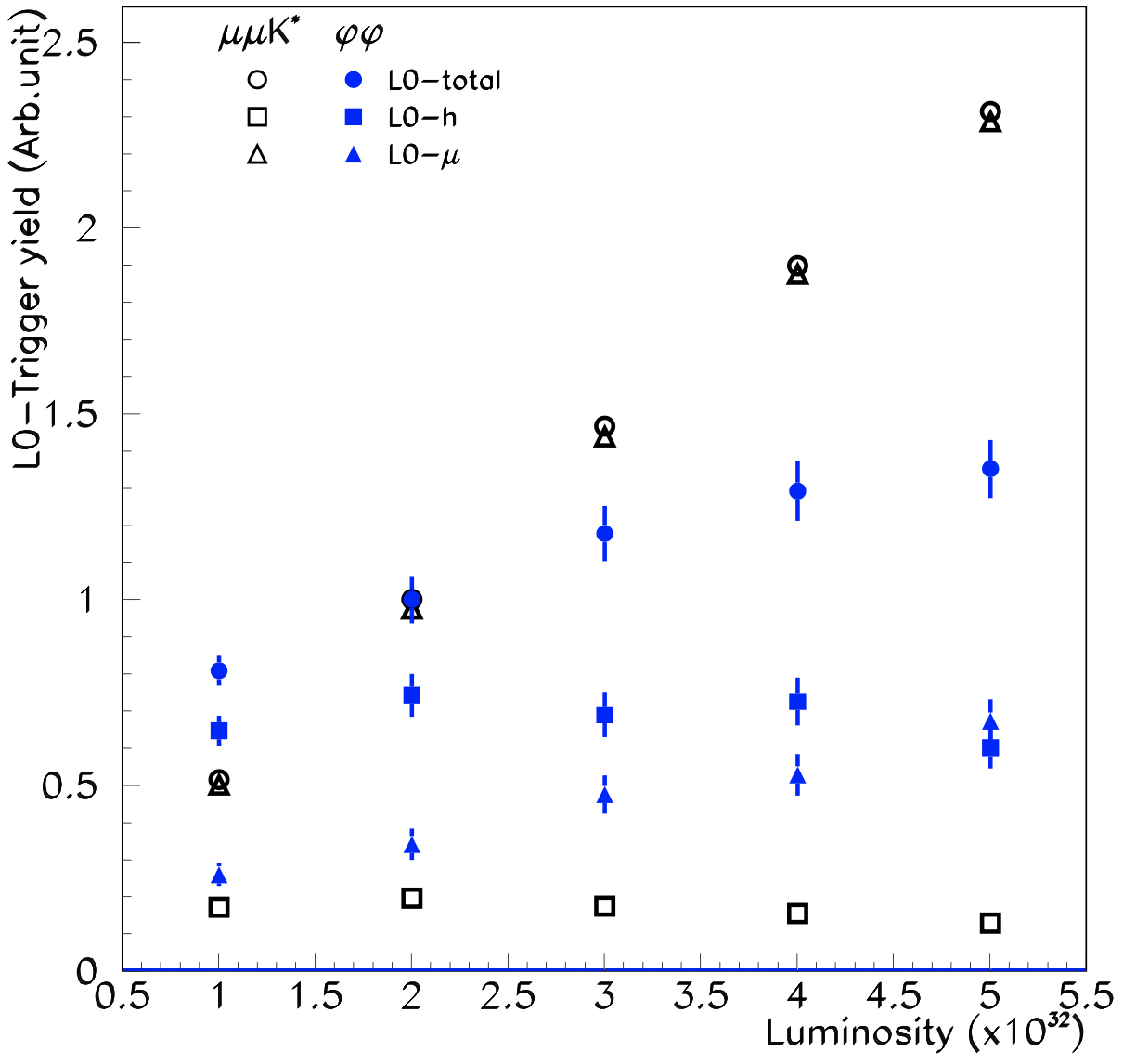}
    \caption{Yield of LHCb's first-level fixed-latency trigger as a function of luminosity for two decay channels: $B^0\to \mu\mu K^{*0}$ (open points) and $B^0_s\to \phi\phi$ (closed points). The total trigger yield, and the contributions from 
    hadron calorimeter and muon triggers are shown separately. As can be seen, for the hadronic channel the trigger yield rapidly 
    plateaus above around $4\cdot 10^{32}$~cm$^{-2}$s$^{-1}$. Reproduced from~\cite{Nakada:1100545}. }
    \label{fig:lhcbtrigyieldvsl0}
\end{figure}

The LHCb experiment was designed as a forward spectrometer optimised for the study of beauty hadron decays. 
Compared to the general QCD processes which occur in a proton-proton collision, beauty hadrons are heavier and have a relatively long lifetime. 
The distinguishing characteristic (signature) of a beauty hadron decay is therefore a set of of charged tracks which have significant momentum transverse to the beamline and originate from a decay vertex displaced from the primary proton-proton collision point.
This combination had already been exploited by CDF's tracking trigger, but the first 
LHCb detector could not read out its tracker at the full LHC bunch crossing rate of $\sim\! 25$~MHz but  only at 1~MHz. 
This is a similar constraint to that which UA1 faced almost thirty years earlier, and LHCb's trigger unsurprisingly 
followed~\cite{Antunes-Nobrega:630828} a conceptually similar architecture. 
The first level LHCb trigger was based on a fixed-latency reconstruction of muon stubs and calorimeter clusters using custom FPGA boards. 
This design led to a bottleneck for calorimeter triggers at instantaneous luminosities above around $4\cdot 10^{32}$~cm$^{-2}$s$^{-1}$, as shown in Figure~\ref{fig:lhcbtrigyieldvsl0}, a fact which was known to the collaboration before datataking began and which motivated the design of the upgraded LHCb detector as we will discuss in the next section.
It was nevertheless the best which could be achieved with the electronics available at the time.

For its second-level triggers, LHCb considered both a two-level approach akin to ATLAS and a one-level approach 
akin to CMS. The former would have been based around a dedicated FPGA architecture searching for displaced 
vertices in the vertex detector and associating them with hits in a tracker station close to the LHCb magnet in order 
to allow a rough momentum estimate; events passing this second-level track trigger would then have been made available 
to a fully software third-level trigger implemented on a farm of CPU processors. After extensive debate LHCb chose
to forgo this second trigger level and bet that the computing technology available when the detector started datataking 
would allow all events accepted by the first-level trigger to be processed in software. A side effect of the use of a 
second-level trigger based on commodity CPU processors was that these processors were available to LHCb for producing 
simulated events during periods where there was no datataking. This very substantially increased the peak computing 
power available to LHCb for simulation, almost doubling it during certain periods of time.
%a fact which the collaboration has made very extensive use of (and which again echoes UA1).
%\textcolor{red}{I think this needs a bit of explanation of rewording}
% I tried, see what you think

While discarding the idea of a custom FPGA track reconstruction, LHCb still kept the idea of splitting this all-software 
trigger into two stages. The first stage was meant to ``confirm'' the decision of the fixed-latency first-level trigger by 
associating tracks and displaced vertices with first-level trigger objects while the second stage would then perform a 
nearly full, offline-like, event reconstruction. This again echoed the way that UA1's third-level trigger used drift chamber 
information to confirm the muon tracks and calorimeter clusters/jets identified by the first- and second-level triggers, as well 
as the ATLAS trigger design. 
At LHCb, however, this approach faced the conceptual difficulty that LHCb's fixed-latency trigger was only around~\cite{LHCb:2018zdd} 
40-50\% efficient for beauty particles which decayed into hadrons or electrons. And the fixed-latency trigger was far less 
efficient for charm hadrons, which grew in importance to LHCb's physics programme as datataking went on. Beauty and charm 
hadrons are produced in pairs at the LHC and their acceptance is also highly correlated: if one beauty or charm hadron 
produced in a collision is within the LHCb acceptance, the other hadron of the same flavour is also highly likely to be within the acceptance. 
This meant that around one half of hadronic beauty decays, and a larger fraction of charm 
decays, passed the fixed-latency trigger not because of the signal of interest to offline analysis but because of the other 
beauty or charm hadron produced in the same bunch crossing. These events would be rejected by the confirmation strategy in the 
2nd level trigger despite being perfectly usable for physics analysis. 

This led to a further simplification~\cite{Gligorov:2011zya,Aaij:2012me} of the LHCb trigger, in which the first software 
stage principally searched for a single track with significant transverse momentum and displacement from all proton proton 
collisions in the event, augmented by additional triggers which increased efficiency for muon and dimuon final states. 
The use of a single displaced track rather than the two-track vertex (as had been done at CDF) was motivated by the fact 
that it was computationally cheaper to reconstruct very high momentum, and therefore straight, tracks. At the same time, 
almost all beauty hadron decays of interest to LHCb produced at least one such track, typically with transverse momentum 
above around $1.25$~GeV. As the computing power available to LHCb's variable-latency trigger grew, displaced vertex 
triggers were reintroduced to improve efficiencies for charm and other softer hadrons which did not produce a 
single very high momentum track as often. The single track trigger remained a mainstay of LHCb's real-time selection 
until the recent upgrade, however. This was not only because of its excellent performance for beauty hadron decays 
but also because of its robustness against data-simulation differences and the ease with which its performance could 
be calibrated and emulated offline. 

\section{From real-time selection to real-time analysis}
\label{sec:rtahllhc}

Particle colliders can increase their instantaneous luminosity by colliding bunches more frequently or by having denser 
collisions with multiple particle interactions (pileup) in each bunch crossing. Conceptually these are two sides of the 
same coin, with an ever-increasing number of interactions separated in either less time or space. 
In practice, the trend has been towards an 
increasing number of interactions per bunch crossing, from a few at the Tevatron, to tens in the first period of LHC 
operations, and to between $140$ and $200$ (and perhaps even higher) for the planned high luminosity LHC (HL-LHC). Since each underlying proton-proton collision is a physically independent process, and traditional trigger systems select entire bunch crossings containing 
a physics signal of interest, this means that signals are embedded in events containing an ever-increasing amount of 
spurious data. This in turn makes the signals more time-consuming to analyse and more expensive to record to permanent 
storage. This problem is being addressed by technological and methodological improvements on two levels: improvements to the 
ability to select events of interest with high purity at the earliest trigger stage, and the development of methods which allow 
high-level analysis objects to be built in real-time, mitigating the need to save the entire detector data 
to permanent storage.

The Phase-2 upgraded CMS L1 trigger~\cite{Zabi:2020gjd} is designed to achieve the same physics acceptance as the original 
CMS detector despite the harsher pile-up environment of the HL-LHC. 
This is achieved by introducing the reconstruction of tracks which, combined with high granularity information from calorimetry 
and muon detectors, allow for a reconstruction of physics objects with a significantly improved resolution. This effectively moves a 
great deal of the functionalities of high-level triggers, currently running on farms of commodity processors, 
closer to the detector into the Level-1 trigger hardware.
The needed processing capabilities are provided by the use of modern FPGAs and fast optical links together with an increase of maximum fixed latency from 4 $\mu$s to 12.5 $\mu$s. 
The total maximum Level-1 trigger rate is increased from 100 kHz to 750 kHz.
Because of the described technological evolution and 
unprecedented computing power, combined with access to information from nearly the whole detector, the upgraded CMS L1 trigger is 
actually expected to further expand the reachable phase-space and allow for an enriched physics program compared to the original CMS, as summarized in Table \ref{tab:CMSL1ExtendedMenu}.
For example, full analysis-like algorithms selecting events based on correlated topological information from the full detector 
have been shown to run in the FPGAs. 
These include precise reconstruction of low mass resonances in searches for signatures 
with very soft muons or tracks that would otherwise be missed by high pT trigger thresholds, for example $\tau\rightarrow 3\mu$ or 
$B^0_s\rightarrow \phi \phi$. Similarily, triggering on tracks, muons, or jets originating from displaced vertices will 
make the CMS detector significantly more sensitive to searches of BSM physics. 
Measurements of vector boson fusion productions 
of Higgs boson decaying to b-quarks or particles invisible by the detector will be improved by successfully detected in the 
global trigger using machine learning techniques.

\begin{table}[th] 
\caption{Extending physics reach of Phase-2 Upgraded CMS Level-1 Trigger by 7 novel trigger paths that were not possible in Phase-1 (summarized from ~\cite{Zabi:2020gjd}). In the first column, targeted physics topics are listed. In the second column, 
the triggers are listed. The names refer to the type of objects used in reconstruction. In the third column, the online $p_{\textrm{T}}$ and $E_{\textrm{T}}$ 
thresholds are reported, except for the case of the "Single $e$/$\gamma$ beyond tracker $\eta$" and Displaced Tracker-Jet, where offline thresholds are reported, 
corresponding to 95\% and 90\%(50\%) efficiency, respectively. The forth column provides the approximate 
additional pure rates these triggers add to the standard Run2-like menu for the case of HL-LHC pileup 200. 
%the approximate pure rates which they add to the simplified menu. 
}
\label{tab:CMSL1ExtendedMenu} 
\begin{center}
\resizebox{\columnwidth}{!}{
\begin{tabular}	{|l|l|c|c|} 
\hline	       &      &	  Online  & 	Rate  \\
      Physics Target & L1 Triggers &  Threshold(s)    &  $\langle {PU} \rangle = 200$ \\
                 &       & (* for Offline) & \\
	          &       &  [GeV]           &  [kHz]  \\   
\hline
\hline  PDFs, Double Parton Scattering. & Single $e$/$\gamma$ beyond tracker $\eta$   &    36 $^{*}$ &   12       \\[5pt]
\hline  Lepton Flavor Violation: $\tau \rightarrow 3\mu$  & Muon-Jet               &    2, 2, 0.5 &   27    \\[5pt]
\hline  FCNC: $B^0_S\rightarrow\phi(K^+K^-)\phi(K^+K^-)$ & Tracker Tracks $B^0_s$            &    12      &   15    \\[5pt]
\hline  SUSY: Dark photon & Displaced Single (Double) Muon  &    22  (20,15)    &   14 (2)   \\[5pt]
\hline  Exotic Higgs decay via light scalars to hadronic states & Displaced Tracker-Jet $H_{\textrm{T}}$  &   248(153) $^{*}$ &   20  \\[5pt]
\hline  LLPs decaying to hadronic states & Displaced Calo-Jet     &   40        &   20    \\[5pt]
\hline
\hline \multicolumn{3}{|l}{Total rate for above triggers} &  \multicolumn{1}{r|}{110~kHz} 			\\[5pt]
\hline 
\end{tabular}
}
\end{center}
\end{table}

The Upgraded CMS HLT~\cite{CMS:2021kat} is designed to further reduce the accept rate to 7.5 kHz from the previous 1 kHz while increasing the processing power by a factor of 20 using heterogenous computing (CPUs, GPUs, FPGAs), and storage throughput by a factor of 25 to 61 GB/s. This HLT rejection power is obtained by additional improvements in reconstructed object purity by including high granularity information of the pixel tracker detector and a newly introduced timing detector which helps discriminate against fake signatures or detector signals originating from pileup interactions.

While aiming at the same set of physics objectives (and improvements compared to the original and Phase-1 upgraded detectors) as CMS, the ATLAS collaboration chose a different approach for their Phase-2 upgraded trigger system. In particular ATLAS will not introduce tracker trigger reconstruction in their hardware Level-1 trigger. Instead, ATLAS will rely on the reconstruction of tracks in the HLT where the full detector information is available for object reconstruction~\cite{ATLAS:2137107}. The maximum output rate of the ATLAS Level-1 trigger will, however, 
be increased to 1~MHz. Similarly, the maximum allowed latency will be increased to 10 $\mu$s at Level-1, allowing for significantly more complex algorithms and selections compared to the previous ATLAS trigger systems. This higher Level-1 output rate will necessitate a higher rejection power of the HLT, which is planned to be achieved by the use of heterogeneus computing technology.  

Despite such improvements in triggering techniques, which allow signal to background ratios to be maintained or even slightly 
improved despite the harsher enviroments, for particularly abundant signals, or signals with backgrounds which are irreducible 
even offline, the cost of having to record 
all detector data for each selected bunch crossing can nevertheless impose a fundamental limitation on the trigger efficiency.
This observation has led~\cite{Khachatryan:2016ecr,Aaij:2016rxn,ATL-DAQ-PUB-2017-003,ATLAS:2018qto,ALICEO2} all LHC experiments to 
introduce the concept of a trigger which compresses the data within a selected bunch crossing. The broad goal of all 
these approaches is to not only identify interesting physics signals in real time, as a normal trigger, but to also 
classify the rest of the information collected by the detector according to whether it is associated with this signal 
or not. One frequently used criterion is to save only those reconstructed particles which are associated with the same 
primary interaction as the signal, effectively suppressing pileup at the trigger level. Only the signal and any associated 
information is then saved to long-term storage, reducing the size of each recorded bunch crossing by an order of magnitude 
or more and thereby allowing the trigger output rate to be correspondingly greater. This concept is referred to 
as real-time analysis on LHCb, scouting on CMS, and trigger-level analysis on ATLAS. ALICE does not have a specific 
name for it since ALICE operates in a luminosity regime with only a single heavy ion interaction per bunch crossing. 
However the entire processing model of the ALICE upgrade is based on a real-time compression of the TPC data which 
serves the same purpose: to retain only the data of interest to physics analysis within a given bunch crossing and 
therefore be able to record more bunch crossings. In what follows we will refer to it as real-time analysis for brevity.

\begin{figure}[t]
    \centering
    \includegraphics[width=0.7\textwidth]{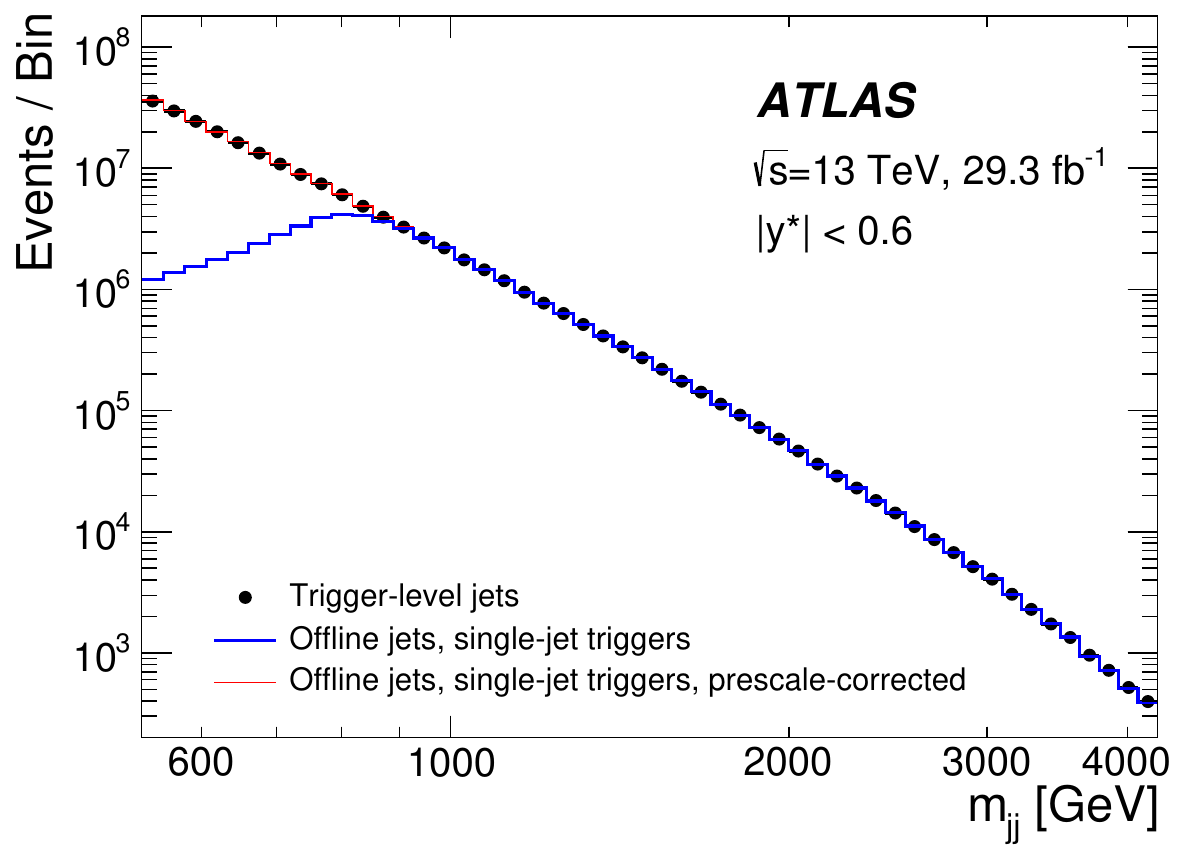}
    \caption{Comparison between the number of dijet events in the data used by the ATLAS real-time analysis (black points), 
    the number of events selected by any single-jet trigger (thicker, blue line), and the events selected by single-jet 
    triggers but corrected for the trigger prescale factors (thinner, red line) as a function of the dijet invariant mass. 
    The definition of $y^*$ is $(y_1-y_2)/2$, where $y_1$ and $y_2$ are the rapidities of the highest- and 
    second-highest-transverse-momentum jets. Figure and caption reproduced from~\cite{ATLAS:2018qto}. }
    \label{fig:atlastla}
\end{figure}

\begin{figure}[t]
    \centering
    \includegraphics[width=0.7\textwidth]{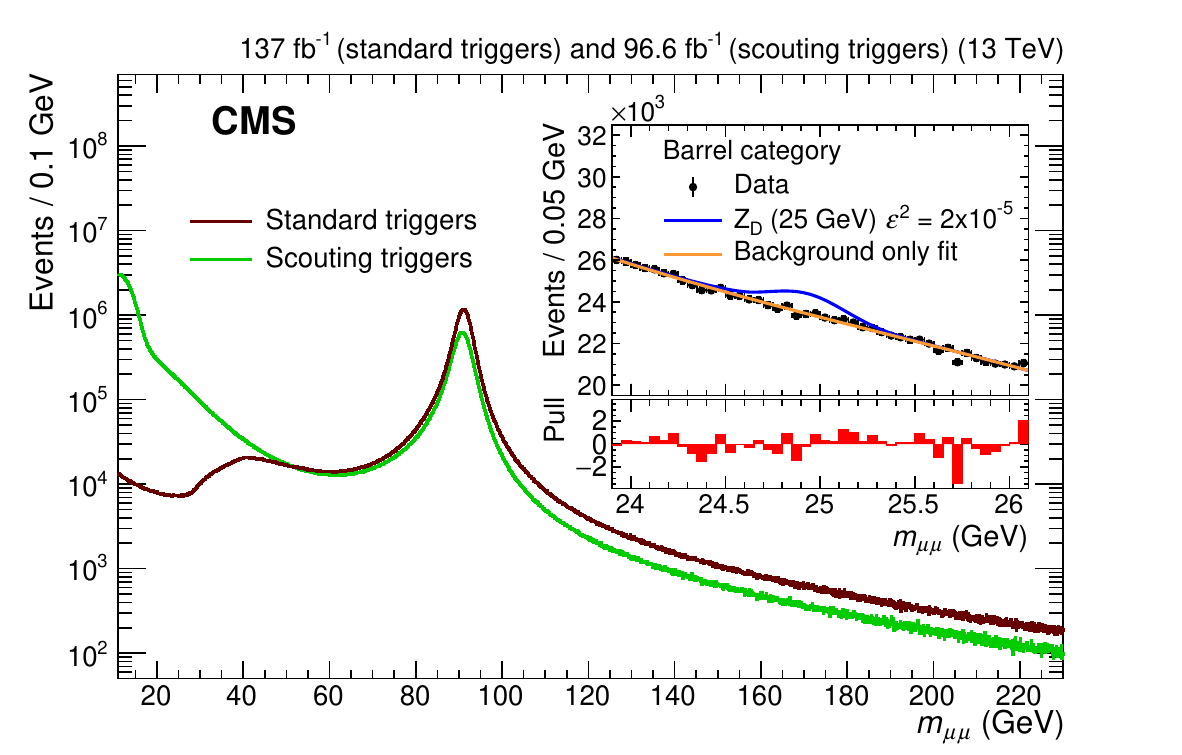}
    \caption{The m$_{\mu\mu}$ distributions of events selected with the standard muon triggers (maroon, darker), and the scouting dimuon triggers (green, lighter). Events are required to pass all the selection requirements. The inset is restricted to events in the barrel category in the mass range 23.9–26.1 GeV. A function describing the background is fit to these data, and a 25 GeV Z$_D$ signal corresponding to $\epsilon^2 = 2 \times 10^{-5}$ is added. The bottom panel of the inset shows the bin-by-bin difference between the number of events in data and the prediction from the background fit, divided by the statistical uncertainty. Figure and caption reproduced from~\cite{CMS:2020dimuscout}  }
    \label{fig:cmsdimuonsout}
\end{figure}

The ATLAS and CMS real-time systems arose from a desire to efficiently search for beyond Standard Model processes in 
situations where the data rate is dominated by irreducible Standard Model backgrounds. If such backgrounds cannot be 
removed even in the final analysis, using the full detector information, then clearly tightening kinematic thresholds in 
order to reduce the associated data rate at the trigger level can only lead to a loss of signal and therefore sensitivity. 
If, however, the information required by the final physics analysis can be reconstructed already at the trigger level, 
then it is possible to save only this information to permanent storage. Since the information required by the final physics 
analysis is typically a fraction of the overall detector data for each event, this allows a much greater number of events 
to be saved for the same total data rate. This is illustrated in Figure~\ref{fig:atlastla} using jet triggers with the ATLAS detector, and in Figure~\ref{fig:cmsdimuonsout} using muon triggers with the CMS detector. In both cases the physics analyses were performed using objects reconstructed in HLT stages.

The logical endpoint of this approach is the CMS trigger scouting system currently under construction for the HL-LHC period, designed to piggy-back on the future Level-1 trigger, also under construction. While the 
upgraded CMS detector cannot be entirely read out at the full LHC bunch crossing rate, all of the detector except the innermost 
tracking layers can be partially read out. This typically means reading out a reduced-granularity version of the detector information, 
or reading out pre-build primitives such as the track stubs created as part of the L1 track trigger described earlier in this review. 
This information is then fed to Level-1 trigger, a set of custom electronics boards equipped with powerful FPGA processors which are able to aggregate and 
combine information from a single detector subsystem or from the different subsystems to construct high-level objects for real-time selection or, as it turns out, also usable by physics analysis. 
The benefits of this L1T scouting system are many fold. It will create opportunities for real-time trigger component diagnostics by capturing their inputs and outputs and will provide invaluable monitoring functionality for almost all stages of selection in Level-1 trigger.  For example, high statistics unbiased data sets collected on this scouting data stream will allow to very quickly identify transient detector problems manifesting in regional triggers, or similarly detect peculiar cases of data loss by re-running algorithms otherwise difficult offline. Equally important, selecting and reconstructing physics objects or specific signal signatures at full collision rate without limitations or biases otherwise imposed by the DAQ and the trigger will allow for an easier luminosity measurement.  

Although the resolution and purity of these objects is of course not equivalent to that achieved in events where the full detector information is available, they are good enough to enable a wide physics program to be carried out directly on the scouting data stream, that is to say at the full LHC bunch crossing rate.  This approach is enabled not only by the computational power of modern FPGA processors, whose memory capacity in particular is approaching levels more typically associated with GPUs and CPUs, but also by advances in high-level programming languages for FPGAs~\cite{7368920,Marc-Andre:2018grr,Marjanovic:2019tle,8356004} which allow 
physicists with relatively limited technical skills to contribute and develop algorithms for the system. 
There has also been a significant effort~\cite{Duarte:2018ite} in the wider community, not only in the context of scouting 
but for all fixed-latency triggers, to enable the efficient use of 
machine learning algorithms and neural networks in FPGA processors. This too brings the types of analyses which can be performed in 
real time using the scouting information closer to what physicists would traditionally have been able to accomplish with data which is 
recorded to permanent storage.  

\begin{figure}[t]
    \centering
    \includegraphics[width=0.7\textwidth]{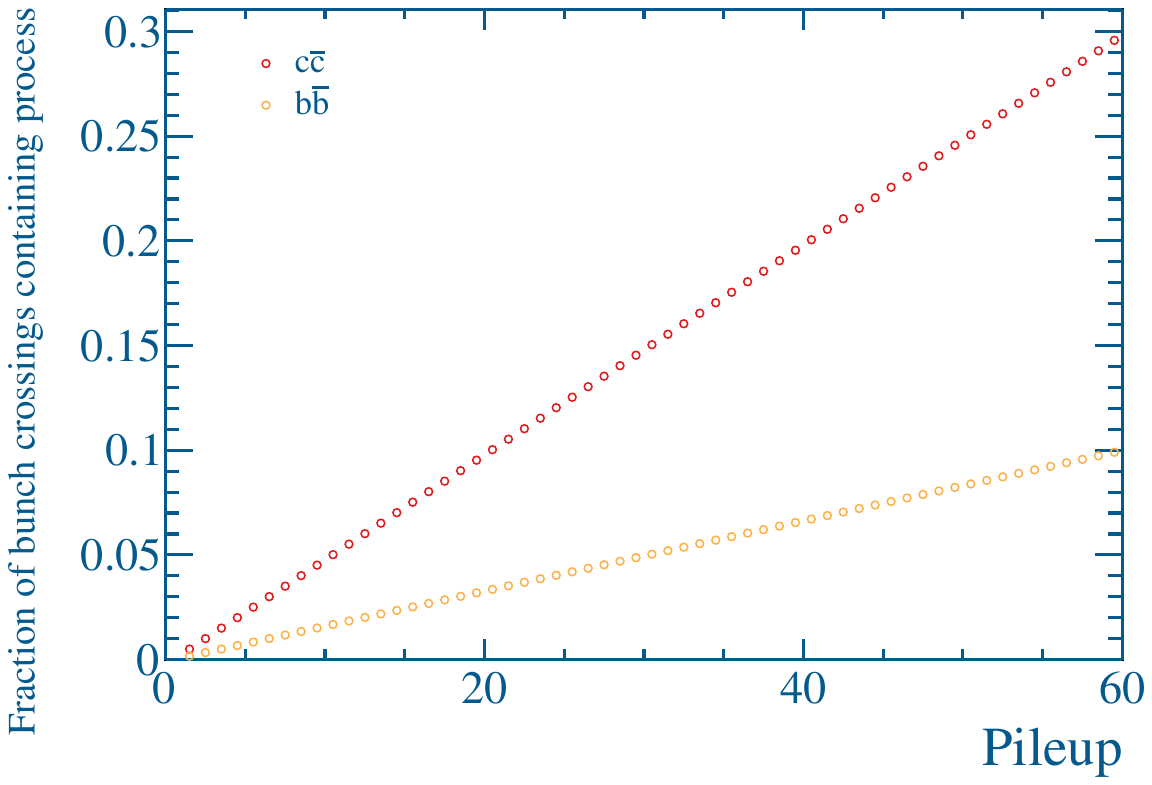}
    \caption{The fraction of LHC bunch crossings which produce a $b\bar{b}$ or $c\bar{c}$ quark pair that can be partially reconstructed in the 
    acceptance of the LHCb detector as a function of the pileup. Based on studies originally documented in Ref.~\cite{Fitzpatrick:1670985}.}
    \label{fig:lhcbbbarvspileup}
\end{figure}

The physics motivation for real-time analysis in LHCb\footnote{Refs.~\cite{Fitzpatrick:1670985,Aaij:2016rxn,Gligorov:2018fuk,Aaij:2019uij} 
contain more detailed and technical discussions of this topic.} is summarized in Figure~\ref{fig:lhcbbbarvspileup} which shows the fraction of 
LHC bunch crossings which produce a $b\bar{b}$ or $c\bar{c}$ quark pair that can be partially reconstructed in the acceptance 
of the LHCb detector as a function of the pileup. The definition of partially reconstructed follows~\cite{Fitzpatrick:1670985} 
and requires a two-track vertex in the LHCb vertex detector as well as at least 2~GeV of momentum transverse to the beamline 
and a decay time greater than $0.2$~picoseconds. These values are chosen because those are the typical minimum transverse 
momentum and decay time thresholds used for signal selection in LHCb physics analyses. We can then see that during LHC 
Runs 1~and~2, when LHCb operated at a pileup of around 1, LHCb's trigger could rely on a combination of transverse momentum 
and displacement from the primary interaction, analogously to CDF and HII. In Run~3 however LHCb is designed to operate at a 
pileup of 6 and this strategy begins to break down. While it would still be just about possible to select beauty hadrons 
in this way, the rate of selected charm would exceed any reasonable trigger rate. And looking towards Runs~5~and~6, when LHCb 
aims to operate at a pileup of between 30 and 40, the approach breaks down even for beauty hadrons. 

There are three fundamental constraints on the allowed output rate of a trigger: the available permanent storage space, the 
ability of physics analysts to access data saved to permanent storage, and the available computing resources to fully reconstruct 
the data saved to permanent storage with the best detector performance and resolutions. The third constraint can be removed by 
aligning and calibrating the detector in near-real-time, as LHCb has been doing since 2015, and subsequently executing the ultimate 
fidelity detector reconstruction in the trigger itself. The first two constraints are closely connected to each other, since larger 
recorded data sets are by definition more complex for individual physics analysts to access. While ensuring that analysts can 
promptly and reliably access large recorded data sets is an interesting problem in its own right it falls outside the scope of this 
review -- Ref.~\cite{Bird:2011zz} is a good introduction to it. This leaves the size of the recorded dataset as a constraint, 
where the number of events selected by the trigger and the size of each event are free parameters. 

\begin{figure}[t]
    \centering
    \includegraphics[width=0.7\textwidth]{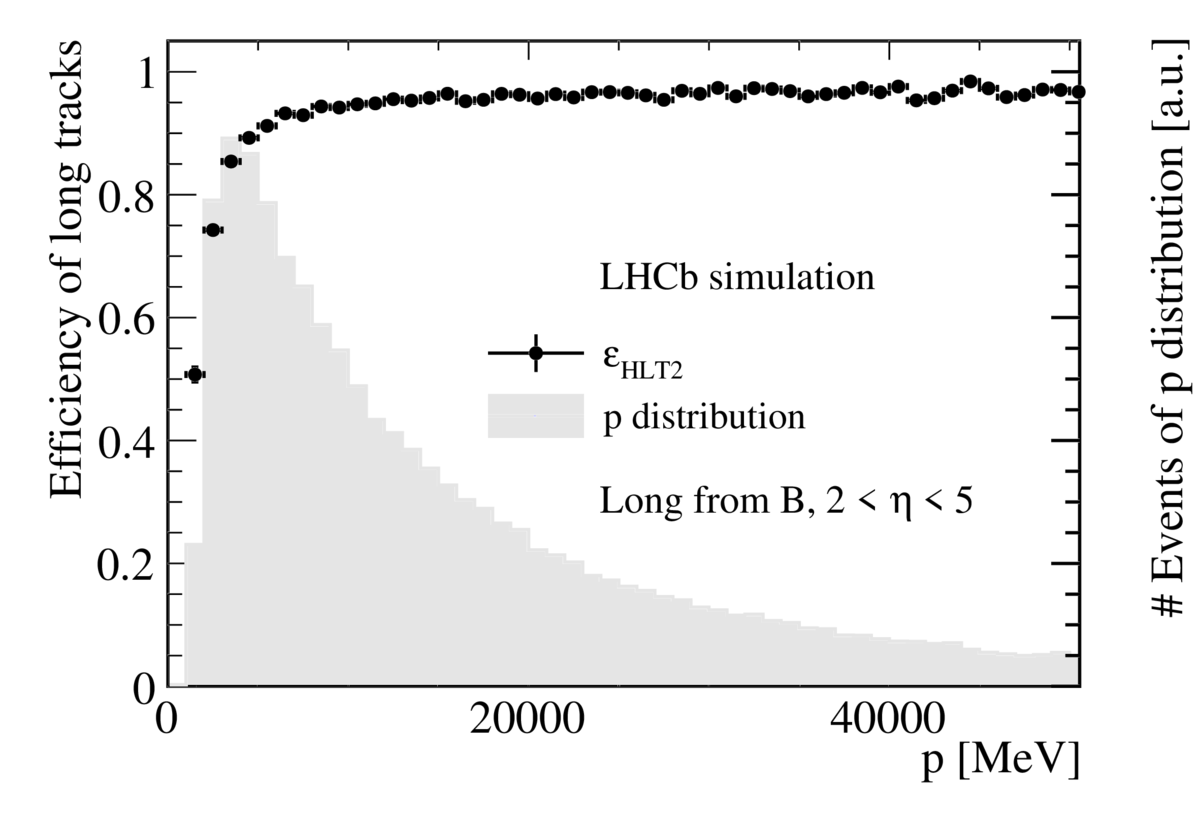}
    \caption{The best achievable efficiency to reconstruct ``long'' tracks, which traverse the entire LHCb tracking system, as a 
    function of track momentum (solid points) compared to the distribution of such tracks in $B$ meson decays in the pseudorapidity 
    range $2<\eta <5$ (shaded histogram). As can be seen the signal populates the area where the detector efficiency varies rapidly 
    as a function of the kinematics. Since this is the efficiency for one single track the overall impact of this dependence on the 
    signal efficiency as a whole is often quadratic or cubic. Figure reproduced from Ref.~\cite{LHCB-FIGURE-2021-003}.}
    \label{fig:lhcbeffturnon}
\end{figure}

Since the rate of bunch crossings containing charm or beauty hadrons saturates the trigger bandwidth -- several times over if 
projecting to Runs 5~and~6 -- it is not possible to reduce the rate of selected events without throwing away a significant 
fraction of signals which are potentially interesting for physics analysis. Moreover, a peculiarity of LHCb emerges when 
comparing Figure~\ref{fig:lhcbeffturnon} to the ideal case illustrated in Figure~\ref{fig:turn-on-curves}. 
In the ideal case the trigger turn-on curve is induced by imperfections or simplifications in the real-time reconstruction 
and selection, and the goal is to make this curve as close as possible to a step function. In the LHCb case the signal of 
interest populates the region where the reconstruction efficiency varies steeply as a function of kinematics, not because 
of simplifications in the real-time algorithms but because of the detector layout and design. It follows that many LHCb 
physics analyses have to efficiency-correct (unfold) the observed signal yields as a function of the signal kinematics. 
This is achieved by using data calibration samples to tune the detector simulation, in particular charm and charmonia 
decays which can be cleanly selected in data using minimally biasing requirements. Because of this, even if you restricted 
LHCb's physics program to a subset of rarer beauty and charm hadron decays which do not saturate the trigger bandwidth, 
the required calibration samples would still push the bandwidth to exceed available resources. This leaves only reducing 
the size of each event on disk as a solution, which is what LHCb implemented in 2015 and has incrementally upgraded since. 

\begin{figure}[t]
    \centering
    \includegraphics[width=0.7\textwidth]{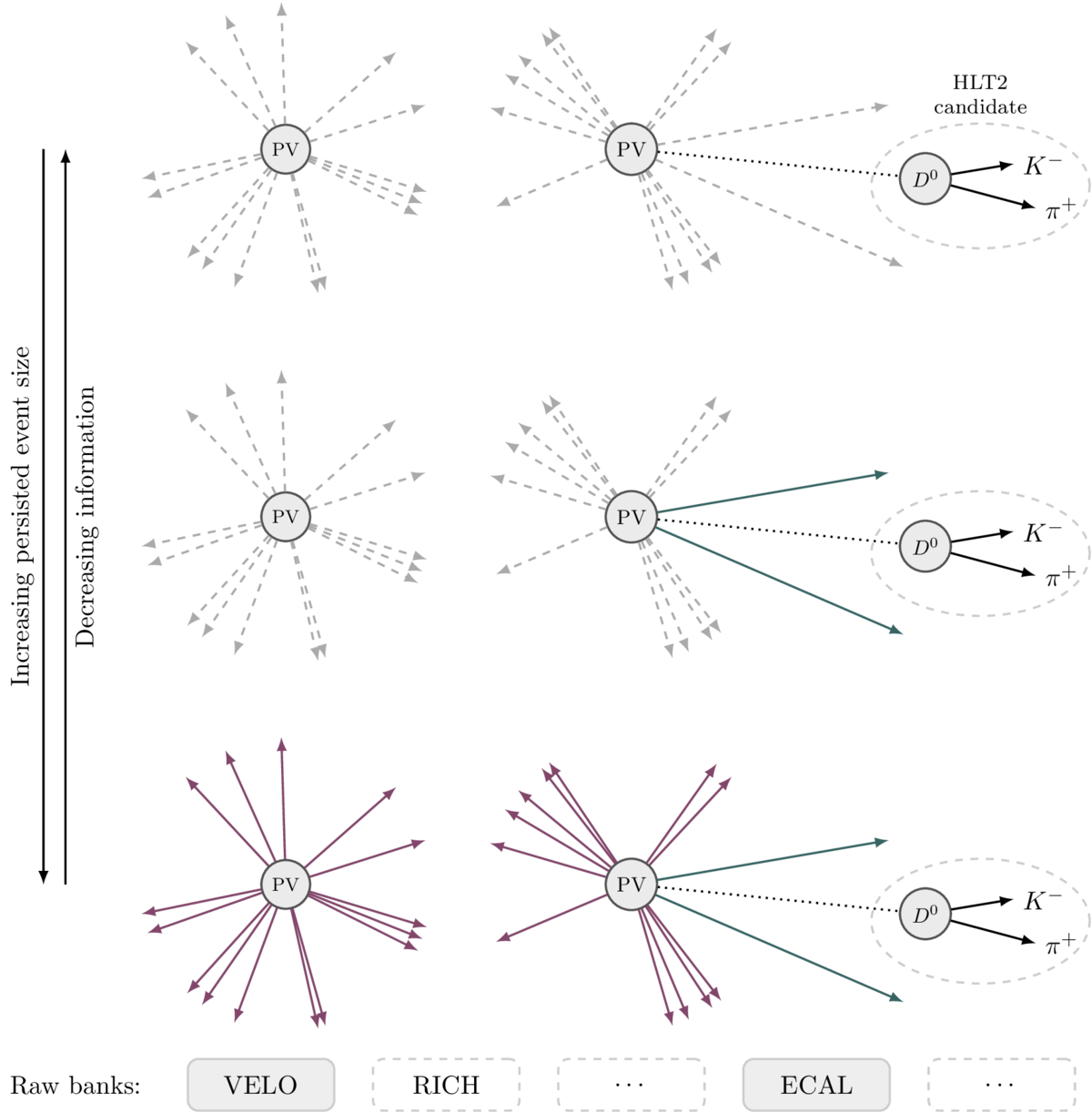}
    \caption{A cartoon of the same reconstructed event with varying levels of object persistence: only the analysis-level signal 
    candidate (top); the signal candidate as well as information about the $pp$ collision in which it was produced (middle); all 
    reconstructed physics objects (bottom). Solid objects are those persisted in each case. A trigger selection may also ask for 
    one or more sub-detector raw detector data ``banks'' to also be stored, shown as solid rectangles. 
    Figure reproduced from Ref.~\cite{Aaij:2019uij}.}
    \label{fig:selectivepersist}
\end{figure}

The most important ingredient of this event size reduction is the full analysis-quality detector reconstruction performed 
in LHCb's second level software trigger. This means that the entire object of interest to a physics analysis -- typically 
a particular beauty or charm hadron decay -- can be reconstructed and associated to a primary interaction with the same 
performance as in the offline physics analysis. The simplest and most economical way of reducing event size is to then save 
only this analysis-level object together with some information about the primary interaction associated with it. It is 
sometimes desirable to store further information, for example about other particles produced in the same primary 
interaction. Example use cases for these particles include tagging the initial quark flavour content of the analysis 
object~\cite{LHCb:2012dgy} or determining if the analysis object is isolated from the rest of the primary interaction associated 
with it. The most powerful use of this ``selective persistance'' is however~\cite{Aaij:2019uij} to allow only part of the analysis 
object to be reconstructed in the trigger, while storing all particles which may eventually be combined with it. This 
kind of trigger-level pileup suppression makes events selected by the trigger to look identical to the physics analyst 
no matter what pileup the detector was operating at. Finally it is possible to store information about all reconstructed 
physics objects in the event while discarding the raw detector data (hits). These three approaches are illustrated in 
Figure~\ref{fig:selectivepersist}.

The relative reduction in event size ranges from around one order of magnitude 
when only saving the analysis-level object to around a factor two when only discarding the raw detector data. LHCb 
deploys several thousand distinct selections in its second-level trigger to satisfy all physics analysis needs, and 
each of these selections would ideally be able to request the optimum amount of event information for the corresponding 
analysis. In practice selections with common physics objectives are grouped together to minimize the number of events 
which physics analysts have to access. For example, an analysis looking for the forbidden lepton-flavour violating 
decay $\tau\to 3\mu$ would not want to sift through all the events used to study $CP$ violation in $D^\pm\to K^+K^-\pi^\pm$ 
decays. 

The ALICE processing model for Run~3 and beyond, known as $O^2$, is driven~\cite{ALICEO2} by three fundamental constraints: the 
fact that ALICE's time projection chamber is around 90\% of the overall event size in lead-lead collisions, the fact 
that lead-lead collisions occurring at 50~kHz drive the overall data volume of the experiment, and the fact that there is 
no way to efficiently select a small subset of interesting lead-lead collisions for physics analysis. Therefore the 
real-time processing focuses on compressing the TPC data by clustering, finding tracks of interest to all ALICE physics 
analyses, and then removing raw TPC information as well as clusters not associated to the found tracks. This is made 
possible by calibrating and reconstructing the detector in real time, using a combination of synchronous and 
asynchronous processing steps. From a conceptual point of view this system is fairly analogous to LHCb's selective 
persistance, although the technical implementations are very different because of the differences in detector geometry 
and readout architecture. 

\begin{figure}[t]
    \centering
    \includegraphics[width=0.6\textwidth]{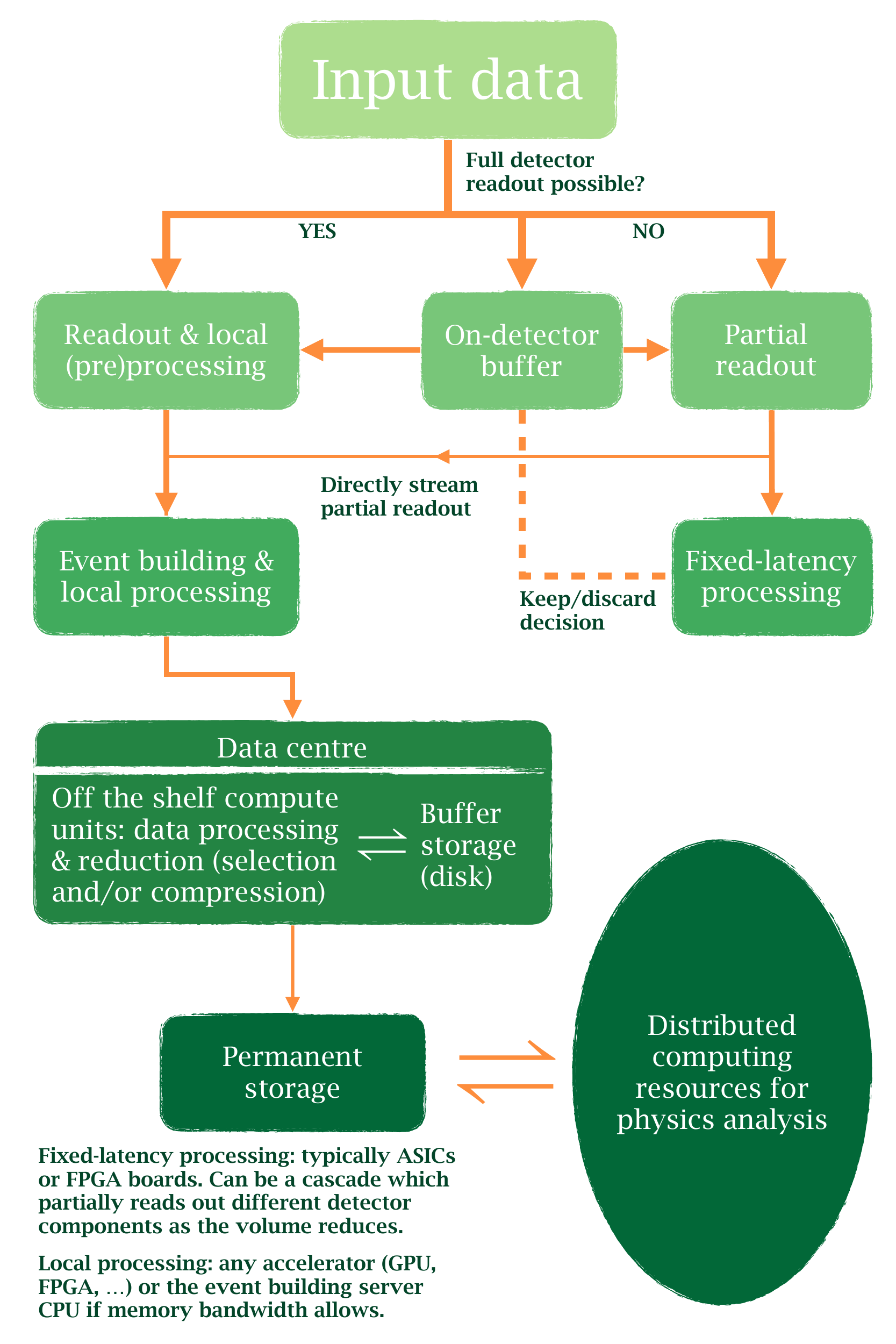}
    \caption{Illustration of the general concepts underpinning contemporary trigger/DAQ architectures.}
    \label{fig:generaltriggerarch}
\end{figure}

Putting all this evolution together, we can sketch a general modern trigger architecture in Figure~\ref{fig:generaltriggerarch}. 
The prejudice of this sketch is that if the whole detector can be read out at the full collision rate then it should be read out 
at the full collision rate, leading to the dataflow on the left of the diagram. If this is too expensive or would insert too 
much material into the detector acceptance, fixed-latency selections based on a partial detector readout should be used. 
Since data rates are typically lower in the outer parts of the detector, these will usually be based on calorimeter or muon 
system information, as has been the case ever since UA1. We also see on the diagram some of the innovations from the LHC era, 
notably the concept of directly streaming the partial detector readout for later physics analysis as in the CMS scouting system. 
Notice that at every stage, whether when reading out the data or building events, there is some room for local processing or 
pre-processing of the data. Examples include transforming detector hits into clusters, neighbouring clusters into track stubs 
and primitives, or selecting events using accelerator boards in the event building servers as in the LHCb Run~3 system.

\section{Impact of commercial computing trends on real-time architectures}
\label{sec:rtatrends}

An important distinction between data processing, whether real-time or otherwise, and most other detector developments
in High Energy Physics is the extent to which our field relies on modifying or adapting existing commercial data
processing solutions rather than leading the development of new ones. This isn't simply because we don't have the critical
mass to design our own processing hardware, but also because the problems our data processing has to solve do not always 
match the problems the rest of scientific or commercial computing wants to solve. This impacts both the hardware choice 
of processing architectures and the development of software which can best exploit them. In the case of fixed-latency 
processing there is now a broad preference for FPGAs over ASICs, both because of their greater flexibility 
and because the continuing improvements to high-level synthesis tools~\cite{Loncar:2020hqp} make FPGAs more and more accessible to 
non-specialist developers. The situation is more complex in variable-latency processing. 

An illustrative example of the challenges facing variable-latency processing in HEP is the ongoing struggle of 
our field to properly exploit today's generic supercomputers or, as they are commonly known, High Performance Computing 
(HPC) centres. The difficulty isn't that those centres don't have enough computing power for our needs: at the time of 
writing, the world's biggest HPC centre has around one hundred times the computing capacity of LHCb's real-time data 
processing centre. This computing power is however optimised for solving problems which involve relatively small volumes 
of data but a large amount of computation, notably training machine learning models. By contrast, the primary use case 
in our field is processing very large volumes of data with a relatively small computational burden per unit data processed. 
Phrased another way, a typical scientific computing application outside HEP reads a little bit of data into memory and 
then uses it as a basis for a great deal of computing resulting in a different, but again relatively small, output data 
file. By contrast a typical HEP application ingests a steady stream of data, performs computations which transform that 
data, and then writes the transformed data stream out.

To put this into context, the Chat GPT large language model was trained on 570~GB of filtered 
data\footnote{\href{https://gptblogs.com/chatgpt-how-much-data-is-used-in-the-training-process}{https://gptblogs.com/chatgpt-how-much-data-is-used-in-the-training-process}} which is roughly 
equal to the volume of triggered data collected by the LHCb experiment in one minute of nominal running. Efficiently 
processing HEP datasets therefore requires high throughput in the processing centres, something which HPC centres do not 
otherwise need to support. This difficulty is amplified by the computing architectures deployed in these centres. 
In the early 2000s, the vast majority of both HPC and HEP data processing was performed by CPUs. Today however 
most\footnote{\href{https://www.top500.org/}{https://www.top500.org/}} HPC sites operate heterogeneous architectures 
consisting of at least CPU and GPU processors, with the majority 
of computing power delivered by the GPUs. Fully loading a GPU with compute is more complex than fully loading a CPU 
because the available theoretical FLOPS are so much greater: for example, the NVIDIA A5000 GPU has around 20 times the 
theoretical FLOPS\footnote{\href{https://en.wikichip.org/wiki/WikiChip}{https://en.wikichip.org/wiki/WikiChip}} of 
an AMD EPYC 7502 CPU. And a typical data centre server will host eight GPU cards. It is 
therefore inherently difficult for a HEP application which aims to transform data to efficiently exploit these GPUs.

Where experiments can design and operate their own data centres, GPUs offer very real benefits, particularly if entire 
processing steps can be implemented on the GPU so as to minimize the overhead of CPU-GPU data transfers. ALICE have led 
the way among collider experiments in the development and adoption of GPUs for real-time processing, and already in 2015 
showed significant gains from using GPUs rather than CPUs to perform the most compute-intensive parts of their real-time 
processing. In the case of LHCb a like-for-like cost comparison~\cite{LHCb:2021kxm} in 2020 showed a factor $2-4$ gain from deploying 
GPUs even before their better energy efficiency~\cite{Aaij:2021oqw} is taken into account. Moreover the growing amounts of memory 
available on even relatively inexpensive GPUs mean that even HEP applications which do not begin by offloading an entire 
processing step, such as in ALICE, CMS, or NA62, are asymptotically trending towards that endpoint. We can 
therefore anticipate that a significant fraction, if not the majority, 
of HL-LHC real-time processing will be performed on GPUs. Writing high-throughput code for a GPU, or indeed any highly 
parallel architecture, remains a specific skill and one which is not taught in typical physics undergraduate courses. 
In particular the emphasis on memory management as the primary bottleneck, rather than the number of computations carried 
out, is one which requires unlearning a great deal of what students were taught during the decades where the focus was on 
object-oriented programming. On the other hand, for precisely the same reasons which drove the commercial adoption of GPUs 
as the primary HPC architecture, this skill is one which will serve those students well whether they remain in HEP or not. 

There is no reason to believe that the commercial fragmentation of computing architectures will stop with GPUs, in 
particular because power consumption is a crucial component of the total cost of ownership for commercial data centres.
\footnote{In HEP, and at CERN in particular, data centre power consumption remains marginal relative to the accelerator's 
own power consumption. There is nevertheless increasing pressure, which will in any case be amplified by commercial 
trends, to deploy applications which minimize this power consumption.} This is reflected in the increasing prevalence 
of use cases for ARM processors, the use of FPGAs as general-purpose accelerators, as well as the 
development of specialised chipsets such as TPUs and post-GPU architectures for machine learning such as IPUs. 

Using ARM processors requires a comparatively modest investment in software portability, which is now being prioritised across 
the community. As a result they are a real alternative to x86 CPUs and their adoption is mainly a question of how total 
cost of ownership evolves for our applications. In the case of FPGAs there is a strong synergy with their use in HEP 
hardware triggers, and it is plausible to imagine HEP real-time architectures evolving towards a mixture of GPU and 
FPGA accelerators. For other more specialised or novel architectures, however, the barrier of having yet another 
development model which can efficiently exploit their specific balance of memory and compute power is likely to slow adoption. 
After all it took more than a decade from initial R\&D studies on GPUs in HEP to their widespread adoption. Because HEP 
computing is and will remain reactive to commercial pressures, however, unforeseen external revolutions might force the 
field to react more quickly than it might otherwise want to. In particular, there are now a number of initiatives to design 
competitive open source chipsets, for example Power10 and RISC-V. These open the medium-term possibility of a chipset 
whose memory-to-compute balance and instruction set are explicitly optimised for HEP workflows. Whether such a chipset can 
then be economically produced and deployed is an interesting question. However the possibility of inverting the development 
cycle of the last decades, in which HEP adapts its processing model to external hardware, into one where HEP works with other 
areas of scientific computing to help them exploit HEP computational methods and optimised-for-HEP processors is certainly 
tantalising.

\section{Real-time processing beyond the LHC era}
\label{sec:rtafuture}

In this brief review we have attempted to describe the central role which real-time processing has played in collider 
experiments to date, and give a generally educated non-specialist an entry point into this particular sub-domain of 
our field. The end of the HL-LHC will mark the first time in over six decades without an operational high-energy 
hadron collider. Given the clear strategic priority~\cite{EuropeanStrategyGroup:2020pow} given to an $e^+e^-$ 
machine to conduct precision studies of, among other 
things, Higgs boson properties, and the technological challenges associated with a hadron collider which 
would significantly improve on the LHC, many decades may well pass before this changes. That presents difficulties to 
collider physicists specialising in real-time processing, because any $e^+e^-$ machine which succeeds the HL-LHC will 
require only relatively simple real-time processing of its data. Our detector and electronics counterparts can 
profit from the $e^+e^-$ period to push the boundaries of extremely light and granular detectors with advanced timing 
capabilities, and the readout itself is an interesting challenge due to the need to minimize the material which cables 
introduce into the detector acceptance. Once the detectors are read out, however, the data rates are expected to be 
relatively modest, as documented in a recent review~\cite{Brenner:2021mxb}. The real-time processing challenge 
associated with them is consequently also modest. It is therefore likely that the community will either reskill, 
perhaps focusing on detector electronics, or shift to other domains which do have 
emerging real-time challenges such as astrophysics. 

This will pose an interesting challenge to the field if and when a new high-energy collider is built. The design 
for the proposed FCC-hh collider~\cite{FCC:2018vvp}, for example, calls for processing events which are seven times 
more energetic than those of the LHC and also have around seven times more proton collisions per bunch crossing. 
This would be the biggest real-time processing challenge ever attempted by the field, particularly if it retains 
the ambition to use FCC-hh detectors to also conduct precision studies of ``light'' physics rather than simply 
as a discovery machine. Any putative muon collider~\cite{Accettura:2023ked} will face its own real-time challenges 
associated with suppressing the formidable beam backgrounds. Predicting computing technology decades into the future 
is a thankless task, but it is not obvious that technology evolution alone will solve these challenges for us, or 
indeed that commercial computing technology will evolve in a direction useful for our problems in the first place. 
While the quality of real-time processing has rarely made the difference between experiments working 
or not working outright, it has frequently been the determining factor in whether those experiments were able to reach 
their full physics potential. Reconstituting a community to face these challenges after what is likely to be a lengthy 
gap will therefore require care and foresight.

The uncertainty around the long-term cannot, however, diminish the clear and exciting challenges facing real-time 
processing over the next twenty years. The approved and proposed HL-LHC detectors will seek to collect data rates an 
order of magnitude beyond any previous collider experiment, using real-time systems of unprecedented complexity and 
flexibility. All three major computing architectures will be exploited, with a near-ubiquitous deployment of machine 
learning models throughout the processing pipelines. Delivering, reliably operating, and evolving these real-time 
systems so as to enable the HL-LHC detectors to fulfill their full physics potential will requires us to develop 
methods and tools which will undoubtedly benefit scientific domains beyond HEP for decades to come.

\section*{Data availability statement}
There is no data associated to this manuscript.

%% file: template_arxiv.bbl
\begin{thebibliography}{57}
\providecommand{\natexlab}[1]{#1}
\providecommand{\url}[1]{\texttt{#1}}
\expandafter\ifx\csname urlstyle\endcsname\relax
  \providecommand{\doi}[1]{doi: #1}\else
  \providecommand{\doi}{doi: \begingroup \urlstyle{rm}\Url}\fi

\bibitem[Lindenstruth and Kisel(2004)]{LINDENSTRUTH200448}
Volker Lindenstruth and Ivan Kisel.
\newblock Overview of trigger systems.
\newblock \emph{Nuclear Instruments and Methods in Physics Research Section A:
  Accelerators, Spectrometers, Detectors and Associated Equipment},
  535\penalty0 (1):\penalty0 48--56, 2004.
\newblock ISSN 0168-9002.
\newblock \doi{10.1016/j.nima.2004.07.267}.
\newblock URL
  \url{https://www.sciencedirect.com/science/article/pii/S0168900204015748}.
\newblock Proceedings of the 10th International Vienna Conference on
  Instrumentation.

\bibitem[Neufeld(2012)]{6418180}
Niko Neufeld.
\newblock Lhc trigger \& daq - an introductory overview.
\newblock In \emph{2012 18th IEEE-NPSS Real Time Conference}, pages 1--4, 2012.
\newblock \doi{10.1109/RTC.2012.6418180}.

\bibitem[Cittolin(2012)]{Cittolin:2012zz}
Sergio Cittolin.
\newblock {The data acquisition and reduction challenge at the Large Hadron
  Collider}.
\newblock \emph{Phil. Trans. Roy. Soc. Lond. A}, 370:\penalty0 950--964, 2012.
\newblock \doi{10.1098/rsta.2011.0464}.

\bibitem[McCallum(2023)]{prices}
John~C. McCallum.
\newblock {Historical prices of computing equipment}.
\newblock 2023.
\newblock URL \url{https://jcmit.net/index.htm}.
\newblock Historical prices of computing equipment
  \href{https://jcmit.net/index.htm}{https://jcmit.net/index.htm}.

\bibitem[Aaij et~al.(2014)]{Aaij:2014jba}
R.~Aaij et~al.
\newblock {LHCb Detector Performance}.
\newblock \emph{Int. J. Mod. Phys. A}, 30\penalty0 (arXiv:1412.6352.):\penalty0
  1530022. 82 p, Dec 2014.

\bibitem[Aad et~al.(2008)]{Aad:1129811}
G~Aad et~al.
\newblock {The ATLAS Experiment at the CERN Large Hadron Collider}.
\newblock \emph{JINST}, 3:\penalty0 S08003. 437 p, 2008.
\newblock \doi{10.1088/1748-0221/3/08/S08003}.
\newblock URL \url{https://cds.cern.ch/record/1129811}.
\newblock Also published by CERN Geneva in 2010.

\bibitem[Chatrchyan et~al.(2008)]{Chatrchyan:1129810}
S~Chatrchyan et~al.
\newblock {The CMS experiment at the CERN LHC. The Compact Muon Solenoid
  experiment}.
\newblock \emph{JINST}, 3:\penalty0 S08004. 361 p, 2008.
\newblock \doi{10.1088/1748-0221/3/08/S08004}.
\newblock URL \url{https://cds.cern.ch/record/1129810}.
\newblock Also published by CERN Geneva in 2010.

\bibitem[Sirunyan et~al.(2020{\natexlab{a}})]{CMS:2020cmk}
Albert~M Sirunyan et~al.
\newblock {Performance of the CMS Level-1 trigger in proton-proton collisions
  at $\sqrt{s} =$ 13 TeV}.
\newblock \emph{JINST}, 15\penalty0 (10):\penalty0 P10017, 2020{\natexlab{a}}.
\newblock \doi{10.1088/1748-0221/15/10/P10017}.

\bibitem[Cortina~Gil et~al.(2017)]{NA62:2017rwk}
Eduardo Cortina~Gil et~al.
\newblock {The Beam and detector of the NA62 experiment at CERN}.
\newblock \emph{JINST}, 12\penalty0 (05):\penalty0 P05025, 2017.
\newblock \doi{10.1088/1748-0221/12/05/P05025}.

\bibitem[Abi et~al.(2020)]{DUNE:2020lwj}
Babak Abi et~al.
\newblock {Deep Underground Neutrino Experiment (DUNE), Far Detector Technical
  Design Report, Volume I Introduction to DUNE}.
\newblock \emph{JINST}, 15\penalty0 (08):\penalty0 T08008, 2020.
\newblock \doi{10.1088/1748-0221/15/08/T08008}.

\bibitem[Aaij et~al.(2023)]{LHCb:2023hlw}
Roel Aaij et~al.
\newblock {The LHCb upgrade I}.
\newblock 5 2023.
\newblock URL \url{http://cds.cern.ch/record/2859353}.

\bibitem[Ananya et~al.(2014)]{ALICEO2}
Ananya et~al.
\newblock O 2 : A novel combined online and offline computing system for the
  alice experiment after 2018.
\newblock \emph{Journal of Physics: Conference Series}, 513\penalty0
  (1):\penalty0 012037, 2014.
\newblock URL \url{http://stacks.iop.org/1742-6596/513/i=1/a=012037}.

\bibitem[ATLAS(2017)]{CERN-LHCC-2017-020}
ATLAS.
\newblock \emph{{Technical Design Report for the Phase-II Upgrade of the ATLAS
  TDAQ System}}, 2017.
\newblock \doi{10.17181/CERN.2LBB.4IAL}.
\newblock URL \url{https://cds.cern.ch/record/2285584}.
\newblock
  \href{https://cds.cern.ch/record/2285584}{https://cds.cern.ch/record/2285584}.

\bibitem[CMS(2020)]{CERN-LHCC-2020-004}
CMS.
\newblock \emph{{The Phase-2 Upgrade of the CMS Level-1 Trigger}}, 2020.
\newblock URL \url{https://cds.cern.ch/record/2714892}.
\newblock Final version
  \href{https://cds.cern.ch/record/2714892}{https://cds.cern.ch/record/2714892}.

\bibitem[Dam(2022)]{Dam:2021sdj}
Mogens Dam.
\newblock {Challenges for FCC-ee luminosity monitor design}.
\newblock \emph{Eur. Phys. J. Plus}, 137\penalty0 (1):\penalty0 81, 2022.
\newblock \doi{10.1140/epjp/s13360-021-02265-3}.

\bibitem[Dorenbosch(1985)]{Dorenbosch:1985cx}
Jheroen Dorenbosch.
\newblock {Trigger in UA2 and in UA1}.
\newblock \emph{eConf}, C851111:\penalty0 134--151, 1985.

\bibitem[Decamp et~al.(1990)]{decamp:in2p3-00005366}
D.~Decamp et~al.
\newblock {Aleph: a detector for electron-positron annihilations at Lep}.
\newblock \emph{{Nuclear Instruments and Methods in Physics Research Section A:
  Accelerators, Spectrometers, Detectors and Associated Equipment}},
  294:\penalty0 121--178, 1990.
\newblock URL \url{https://hal.in2p3.fr/in2p3-00005366}.

\bibitem[Bocci et~al.(1994)]{Bocci:274052}
V~Bocci et~al.
\newblock {Basic concepts and architectural details of the DELPHI trigger
  system}.
\newblock 1994.
\newblock \doi{10.1109/23.467783}.
\newblock URL \url{https://cds.cern.ch/record/274052}.
\newblock
  \href{https://cds.cern.ch/record/274052}{https://cds.cern.ch/record/274052}.

\bibitem[Abt et~al.(1997)]{H1:1996prr}
I.~Abt et~al.
\newblock {The H1 detector at HERA}.
\newblock \emph{Nucl. Instrum. Meth. A}, 386:\penalty0 310--347, 1997.
\newblock \doi{10.1016/S0168-9002(96)00893-5}.

\bibitem[Baird et~al.(2001)]{Baird:2001xc}
A.~Baird et~al.
\newblock {A Fast high resolution track trigger for the H1 experiment}.
\newblock \emph{IEEE Trans. Nucl. Sci.}, 48:\penalty0 1276--1285, 2001.
\newblock \doi{10.1109/23.958765}.

\bibitem[Amidei et~al.(1988)]{CDF:1987dye}
D.~Amidei et~al.
\newblock {A Two Level Fastbus Based Trigger System for CDF}.
\newblock \emph{Nucl. Instrum. Meth. A}, 269:\penalty0 51--62, 1988.
\newblock \doi{10.1016/0168-9002(88)90861-3}.

\bibitem[Ashmanskas et~al.(2004)]{CDF:2003mka}
Bill Ashmanskas et~al.
\newblock {The CDF silicon vertex trigger}.
\newblock \emph{Nucl. Instrum. Meth. A}, 518:\penalty0 532--536, 2004.
\newblock \doi{10.1016/j.nima.2003.11.078}.

\bibitem[CER(1998)]{CERN-LHCC-98-014}
\emph{{ATLAS level-1 trigger: Technical Design Report}}.
\newblock CERN, Geneva, 1998.
\newblock URL \url{https://cds.cern.ch/record/381429}.

\bibitem[Jenni et~al.(2003)Jenni, Nessi, Nordberg, and Smith]{Jenni:616089}
Peter Jenni, Marzio Nessi, Markus Nordberg, and Kenway Smith.
\newblock \emph{{ATLAS high-level trigger, data-acquisition and controls:
  Technical Design Report}}.
\newblock CERN, Geneva, 2003.
\newblock URL \url{https://cds.cern.ch/record/616089}.

\bibitem[Bayatyan et~al.()]{Bayatyan:706847}
G~L Bayatyan et~al.
\newblock \emph{{CMS TriDAS project: Technical Design Report, Volume 1: The
  Trigger Systems}}.
\newblock URL \url{https://cds.cern.ch/record/706847}.

\bibitem[Cittolin et~al.(2002)Cittolin, Rácz, and Sphicas]{Cittolin:578006}
Sergio Cittolin, Attila Rácz, and Paris Sphicas.
\newblock \emph{{CMS The TriDAS Project: Technical Design Report, Volume 2:
  Data Acquisition and High-Level Trigger. CMS trigger and data-acquisition
  project}}.
\newblock CERN, Geneva, 2002.

\bibitem[Antunes-Nobrega et~al.(2003)]{Antunes-Nobrega:630828}
R~Antunes-Nobrega et~al.
\newblock \emph{{LHCb trigger system: Technical Design Report}}.
\newblock CERN, Geneva, 2003.
\newblock URL \url{https://cds.cern.ch/record/630828}.

\bibitem[Nakada et~al.(2008)Nakada, Ullaland, and Witzelling]{Nakada:1100545}
Tatsuya Nakada, O~Ullaland, and Werner Witzelling.
\newblock {Expression of Interest for an LHCb Upgrade}.
\newblock Technical report, CERN, Geneva, 2008.
\newblock URL \url{https://cds.cern.ch/record/1100545}.

\bibitem[Aaij et~al.(2019{\natexlab{a}})]{LHCb:2018zdd}
Roel Aaij et~al.
\newblock {Design and performance of the LHCb trigger and full real-time
  reconstruction in Run 2 of the LHC}.
\newblock \emph{JINST}, 14\penalty0 (04):\penalty0 P04013, 2019{\natexlab{a}}.
\newblock \doi{10.1088/1748-0221/14/04/P04013}.

\bibitem[Gligorov(2011)]{Gligorov:2011zya}
V.~V. Gligorov.
\newblock {A single track HLT1 trigger}.
\newblock 1 2011.
\newblock URL \url{https://inspirehep.net/literature/928797}.
\newblock
  \href{https://inspirehep.net/literature/928797}{https://inspirehep.net/literature/928797}.

\bibitem[Aaij et~al.(2013)]{Aaij:2012me}
R~Aaij et~al.
\newblock {The LHCb Trigger and its Performance in 2011}.
\newblock \emph{JINST}, 8:\penalty0 P04022, 2013.
\newblock \doi{10.1088/1748-0221/8/04/P04022}.

\bibitem[Zabi et~al.(2020)Zabi, Berryhill, Perez, and Tapper]{Zabi:2020gjd}
Alexandre Zabi, Jeffrey~Wayne Berryhill, Emmanuelle Perez, and Alexander~D.
  Tapper.
\newblock {The Phase-2 Upgrade of the CMS Level-1 Trigger}.
\newblock 2020.

\bibitem[Badaro et~al.(2021)]{CMS:2021kat}
Gilbert Badaro et~al.
\newblock {The Phase-2 Upgrade of the CMS Data Acquisition}.
\newblock \emph{EPJ Web Conf.}, 251:\penalty0 04023, 2021.
\newblock \doi{10.1051/epjconf/202125104023}.

\bibitem[ATL()]{ATLAS:2137107}
{Technical Design Report for the Phase-II Upgrade of the ATLAS TDAQ System}.
\newblock \doi{10.17181/CERN.2LBB.4IAL}.

\bibitem[Khachatryan et~al.(2016)]{Khachatryan:2016ecr}
Vardan Khachatryan et~al.
\newblock {Search for narrow resonances in dijet final states at $\sqrt(s)=$ 8
  TeV with the novel CMS technique of data scouting}.
\newblock \emph{Phys. Rev. Lett.}, 117\penalty0 (3):\penalty0 031802, 2016.
\newblock \doi{10.1103/PhysRevLett.117.031802}.

\bibitem[Aaij et~al.(2016)]{Aaij:2016rxn}
R.~Aaij et~al.
\newblock {Tesla : an application for real-time data analysis in High Energy
  Physics}.
\newblock \emph{Comput. Phys. Commun.}, 208:\penalty0 35--42, 2016.
\newblock \doi{10.1016/j.cpc.2016.07.022}.

\bibitem[ATL(2017)]{ATL-DAQ-PUB-2017-003}
{Trigger-object Level Analysis with the ATLAS detector at the Large Hadron
  Collider: summary and perspectives}.
\newblock Technical Report ATL-DAQ-PUB-2017-003, CERN, Geneva, Dec 2017.
\newblock URL \url{https://cds.cern.ch/record/2295739}.

\bibitem[Aaboud et~al.(2018)]{ATLAS:2018qto}
M.~Aaboud et~al.
\newblock {Search for low-mass dijet resonances using trigger-level jets with
  the ATLAS detector in $pp$ collisions at $\sqrt{s}=13$ TeV}.
\newblock \emph{Phys. Rev. Lett.}, 121\penalty0 (8):\penalty0 081801, 2018.
\newblock \doi{10.1103/PhysRevLett.121.081801}.

\bibitem[Sirunyan et~al.(2020{\natexlab{b}})]{CMS:2020dimuscout}
A.~M. Sirunyan et~al.
\newblock {Search for a Narrow Resonance Lighter than 200 GeV Decaying to a
  Pair of Muons in Proton-Proton Collisions at $\sqrt{s}=13$ TeV}.
\newblock \emph{Phys. Rev. Lett.}, 124\penalty0 (13):\penalty0 131802,
  2020{\natexlab{b}}.
\newblock \doi{10.1103/PhysRevLett.124.131802}.

\bibitem[Nane et~al.(2016)Nane, Sima, Pilato, Choi, Fort, Canis, Chen, Hsiao,
  Brown, Ferrandi, Anderson, and Bertels]{7368920}
Razvan Nane, Vlad-Mihai Sima, Christian Pilato, Jongsok Choi, Blair Fort,
  Andrew Canis, Yu~Ting Chen, Hsuan Hsiao, Stephen Brown, Fabrizio Ferrandi,
  Jason Anderson, and Koen Bertels.
\newblock A survey and evaluation of fpga high-level synthesis tools.
\newblock \emph{IEEE Transactions on Computer-Aided Design of Integrated
  Circuits and Systems}, 35\penalty0 (10):\penalty0 1591--1604, 2016.
\newblock \doi{10.1109/TCAD.2015.2513673}.

\bibitem[Marc-Andr\'e(2018)]{Marc-Andre:2018grr}
T\'etrault Marc-Andr\'e.
\newblock {Two FPGA Case Studies Comparing High Level Synthesis and Manual HDL
  for HEP applications}.
\newblock 6 2018.

\bibitem[Marjanovic(2019)]{Marjanovic:2019tle}
Jan Marjanovic.
\newblock {Low vs High Level Programming for FPGA}.
\newblock In \emph{{7th International Beam Instrumentation Conference}}, page
  THOA01, 2019.
\newblock \doi{10.18429/JACoW-IBIC2018-THOA01}.

\bibitem[Lahti et~al.(2019)Lahti, Sjövall, Vanne, and Hämäläinen]{8356004}
Sakari Lahti, Panu Sjövall, Jarno Vanne, and Timo~D. Hämäläinen.
\newblock Are we there yet? a study on the state of high-level synthesis.
\newblock \emph{IEEE Transactions on Computer-Aided Design of Integrated
  Circuits and Systems}, 38\penalty0 (5):\penalty0 898--911, 2019.
\newblock \doi{10.1109/TCAD.2018.2834439}.

\bibitem[Duarte et~al.(2018)]{Duarte:2018ite}
Javier Duarte et~al.
\newblock {Fast inference of deep neural networks in FPGAs for particle
  physics}.
\newblock \emph{JINST}, 13\penalty0 (07):\penalty0 P07027, 2018.
\newblock \doi{10.1088/1748-0221/13/07/P07027}.

\bibitem[Fitzpatrick and Gligorov(2014)]{Fitzpatrick:1670985}
C~Fitzpatrick and V~V Gligorov.
\newblock {Anatomy of an upgrade event in the upgrade era, and implications for
  the LHCb trigger}.
\newblock LHCb-PUB-2014-027, 2014.

\bibitem[Gligorov(2018)]{Gligorov:2018fuk}
Vladimir~Vava Gligorov.
\newblock \emph{{Conceptualization, implementation, and commissioning of
  real-time analysis in the High Level Trigger of the LHCb experiment}}.
\newblock PhD thesis, Paris U., VI-VII, 2018.

\bibitem[Aaij et~al.(2019{\natexlab{b}})]{Aaij:2019uij}
R.~Aaij et~al.
\newblock {A comprehensive real-time analysis model at the LHCb experiment}.
\newblock \emph{JINST}, 14\penalty0 (04):\penalty0 P04006, 2019{\natexlab{b}}.
\newblock \doi{10.1088/1748-0221/14/04/P04006}.

\bibitem[Bird(2011)]{Bird:2011zz}
Ian Bird.
\newblock {Computing for the Large Hadron Collider}.
\newblock \emph{Ann. Rev. Nucl. Part. Sci.}, 61:\penalty0 99--118, 2011.
\newblock \doi{10.1146/annurev-nucl-102010-130059}.

\bibitem[LHCb(2021)]{LHCB-FIGURE-2021-003}
LHCb.
\newblock {Selected HLT2 reconstruction performance for the LHCb upgrade}.
\newblock 2021.
\newblock URL \url{https://cds.cern.ch/record/2773174}.
\newblock
  \href{https://cds.cern.ch/record/2773174}{https://cds.cern.ch/record/2773174}.

\bibitem[Aaij et~al.(2012)]{LHCb:2012dgy}
R.~Aaij et~al.
\newblock {Opposite-side flavour tagging of B mesons at the LHCb experiment}.
\newblock \emph{Eur. Phys. J. C}, 72:\penalty0 2022, 2012.
\newblock \doi{10.1140/epjc/s10052-012-2022-1}.

\bibitem[Loncar et~al.(2021)]{Loncar:2020hqp}
Vladimir Loncar et~al.
\newblock {Compressing deep neural networks on FPGAs to binary and ternary
  precision with HLS4ML}.
\newblock \emph{Mach. Learn. Sci. Tech.}, 2:\penalty0 015001, 2021.
\newblock \doi{10.1088/2632-2153/aba042}.

\bibitem[Aaij et~al.(2022)]{LHCb:2021kxm}
R.~Aaij et~al.
\newblock {A Comparison of CPU and GPU Implementations for the LHCb Experiment
  Run 3 Trigger}.
\newblock \emph{Comput. Softw. Big Sci.}, 6\penalty0 (1):\penalty0 1, 2022.
\newblock \doi{10.1007/s41781-021-00070-2}.

\bibitem[Aaij et~al.(2021)Aaij, C\'ampora~P\'erez, Colombo, Fitzpatrick,
  Gligorov, Hennequin, Neufeld, Nolte, Schwemmer, and Vom~Bruch]{Aaij:2021oqw}
Roel Aaij, Daniel~Hugo C\'ampora~P\'erez, Tommaso Colombo, Conor Fitzpatrick,
  Vladimir~Vava Gligorov, Arthur Hennequin, Niko Neufeld, Niklas Nolte, Rainer
  Schwemmer, and Dorothea Vom~Bruch.
\newblock {Evolution of the energy efficiency of LHCb\textquoteright{}s
  real-time processing}.
\newblock \emph{EPJ Web Conf.}, 251:\penalty0 04009, 2021.
\newblock \doi{10.1051/epjconf/202125104009}.

\bibitem[Eur(2020)]{EuropeanStrategyGroup:2020pow}
\emph{{2020 Update of the European Strategy for Particle Physics}}.
\newblock CERN Council, Geneva, 2020.
\newblock ISBN 978-92-9083-575-2.
\newblock \doi{10.17181/ESU2020}.

\bibitem[Brenner and Leonidopoulos(2021)]{Brenner:2021mxb}
Richard Brenner and Christos Leonidopoulos.
\newblock {Online computing challenges: detector and read-out requirements}.
\newblock \emph{Eur. Phys. J. Plus}, 136\penalty0 (12):\penalty0 1198, 2021.
\newblock \doi{10.1140/epjp/s13360-021-02155-8}.

\bibitem[Abada et~al.(2019)]{FCC:2018vvp}
A.~Abada et~al.
\newblock {FCC-hh: The Hadron Collider}: {Future Circular Collider Conceptual
  Design Report Volume 3}.
\newblock \emph{Eur. Phys. J. ST}, 228\penalty0 (4):\penalty0 755--1107, 2019.
\newblock \doi{10.1140/epjst/e2019-900087-0}.

\bibitem[Accettura et~al.(2023)]{Accettura:2023ked}
Carlotta Accettura et~al.
\newblock {Towards a Muon Collider}.
\newblock 3 2023.

\end{thebibliography}
